\documentclass[twocolumn]{aastex631}
\usepackage{natbib}
\usepackage{url}
\usepackage{colortbl}
\usepackage{float}
\usepackage{multirow}
\usepackage{graphicx}
\usepackage{lipsum}

\DeclareUnicodeCharacter{2212}{-}

\newcommand{\msun}{\mbox{$\,{\rm M}_\odot$}}
\newcommand{\lsun}{\mbox{$\,{\rm L}_\odot$}}

\newcommand{\oversim}[2]{\protect{\mbox{\lower0.5ex\vbox{%
  \baselineskip=0pt\lineskip=0.2ex
  \ialign{$\mathsurround=0pt #1\hfil##\hfil$\crcr#2\crcr\sim\crcr}}}}}
\DeclareRobustCommand{\VAN}[3]{#2}
\let\VANthebibliography\thebibliography
\def\thebibliography{\DeclareRobustCommand{\VAN}[3]{##3}\VANthebibliography}

\begin{document}

\title{Long-period variable stars in NGC 147 and NGC 185 – II. Their dust production}

%\correspondingauthor{Hedieh Abdollahi}
\email{mahani@ipm.ir}
\correspondingauthor{Atefeh Javadi}
\email{atefeh@ipm.ir}

\author[0009-0002-9957-5818]{Hamidreza Mahani}
\affiliation{School of Astronomy, Institute for Research in Fundamental Sciences (IPM), P.O. Box 1956836613, Tehran, Iran}

\author[0000-0001-8392-6754]{Atefeh Javadi}
\affiliation{School of Astronomy, Institute for Research in Fundamental Sciences (IPM), P.O. Box 1956836613, Tehran, Iran}

\author[0000-0002-1272-3017]{Jacco Th. van Loon}
\affiliation{Lennard-Jones Laboratories, Keele University, ST5 5BG, UK}

\author[0000-0003-2743-8240]{Francisca Kemper}
\affiliation{Institute of Space Sciences (ICE), CSIC, Can Magrans, 08193 Cerdanyola del Vall{\'e}s, Barcelona, Spain}
\affiliation{ICREA, Pg. Llu{\'i}s Companys 23, Barcelona, Spain}
\affiliation{Institut d'Estudis Espacials de Catalunya (IEEC), E-08034 Barcelona, Spain}

\author[0000-0003-1993-2302]{Roya Hamedani Golshan}
\affiliation{Physikalisches Institut der Universit{\"a}t zu K{\"o}ln, Z{\"u}lpicher Str. 77, D-50937 K{\"o}ln, Germany}

\author[0000-0003-0356-0655]{Iain McDonald}
\affiliation{Department of Physical Sciences, The Open University, Walton Hall, Milton Keynes, UK}
\affiliation{Jodrell Bank Centre for Astrophysics, Alan Turing Building, University of Manchester, M13 9PL, UK}

\author[0000-0003-0558-8782
]{Habib G. Khosroshahi}
\affiliation{School of Astronomy, Institute for Research in Fundamental Sciences (IPM), P.O. Box 1956836613, Tehran, Iran}
\affiliation{Iranian National Observatory, Institute for Research in Fundamental Sciences (IPM), Tehran, Iran}

\author[0000-0002-7823-7169] {Hedieh Abdollahi}
\affiliation{School of Astronomy, Institute for Research in Fundamental Sciences (IPM), P.O. Box 1956836613, Tehran, Iran}
\affiliation{Konkoly Observatory, HUN-REN Research Centre for Astronomy and Earth Sciences, MTA Centre of Excellence, Konkoly Thege M. {\'u}t 15-17, Budapest, H–1121, Hungary}

\author[0009-0009-7919-5245]{Sajjad Mahdizadeh}
\affiliation{Physikalisches Institut der Universit{\"a}t zu K{\"o}ln, Z{\"u}lpicher Str. 77, D-50937 K{\"o}ln, Germany}
\affiliation{Department of Physics, Durham University, Odgen Centre For Fundamental Physics West, Lower Mountjoy, South Rd, Durham DH1 3LE, United Kingdom}
%@@@@@@@@@@@@@@@@@@@@@@@@@@@@@@@@@@@@@@@@@@@@@@@@@@@@@@@@@@@@@@@@@@@@@@@@@@@@@@@@@@@@@@@@@@

\begin{abstract}

This study presents a comparative analysis of mass-loss and dust-production rates in the dwarf galaxies NGC 147 and NGC 185, focusing on long-period variables (LPVs) and pulsating asymptotic giant branch (AGB) stars as primary indicators of dust feedback into the interstellar medium. For NGC 147, the total mass-loss rate is calculated as $(9.44 \pm 3.78) \times 10^{-4} \msun yr^{-1}$, with LPV luminosities ranging from $(6.20 \pm 0.25) \times 10^{2} L_\odot$ to $( 7.87 \pm 0.32) \times 10^{3}  L_\odot $. In NGC 185, the total mass-loss rate is higher, at $(1.58 \pm 0.63) \times 10^{-3} \msun yr^{-1}$, with LPV luminosities spanning  $ (5.68 \pm 0.23) \times 10^{2}  L_\odot $ to  $(1.54 \pm 0.66) \times 10^{4}  L_\odot$. A positive correlation is observed between stellar luminosity, intrinsic reddening due to circumstellar dust self-extinction, and elevated mass-loss rates. 
Additionally, comparisons of calculated dust injection rates, two-dimensional dust distribution maps, and observed dust masses provide evidence for a gravitational interaction between NGC 147 and the Andromeda galaxy, which influences the dust distribution within the system.

\end{abstract}


%@@@@@@@@@@@@@@@@@@@@@@@@@@@@@@@@@@@@@@@@@@@@@@@@@@@@@@@@@@@@@@@@@@@@@@@@@@@@@@@@@@@@@@@@@@

\section{Introduction} 

The dynamical evolution and chemical enrichment of galaxies are significantly influenced by the mass-loss and dust production of evolved stars. Due to their substantial material loss during their late evolutionary stages, these stars, which are frequently located on the asymptotic giant branch, are important contributors to the interstellar medium. Investigating the mechanisms driving dust production and understanding how these processes depend on stellar and environmental properties are crucial for shedding light on the broader dynamics of galactic evolution.

Dwarf elliptical (dE) galaxies provide unique environments for studying these phenomena. These galaxies are of considerable interest in modern observational astronomy, because they give vital insights into the interpretation of deep redshift survey results and the evolution of stars and interstellar medium particles \citep{Ferguson94}. The two dE satellites of Andromeda, the most massive component of the Local Group (LG), are NGC 147 and NGC 185. A brief comparison of these galaxies is provided in Table \ref{tab:quantities}. Although they have many similarities, including luminosity \citep{Crnojevic14} and velocity dispersion \citep{Geha10}, they also have significant differences, which are discussed next.

Nothing indicates that star formation has occurred recently in the NGC 147, as it contains no gas and dust. However, those particles have been detected in NGC 185, and its central regions show evidence of recent star formation \citep{Young97,Welch98,Marleau10,DeLooze16}. Moreover, based on the photometry of the red giant branch, NGC 147 has a slightly greater mean metallicity than NGC 185 \citep{Davidge94,McConnachie05,Geha10}. Some contend that NGC 147 and 185 are a bound galaxy pair due to their proximity (58 arc-min) in sky projection \citep{vandenBergh98}. On the other hand, kinematic evidence indicates that they may not be gravitationally bound \citep{Geha10,Watkins13}. While NGC 147 is tidally distorted, NGC 185 is not \citep{Ferguson16}. The star formation history in these two galaxies is another contrast between them. \citet{Hamedani17} demonstrated unique SFH in NGC 147 and 185, confirming that the amount of gas, mass-loss, and dust in their  interstellar medium (ISM) might differ.

\begin{table}
\begin{center}
 \caption{  A  comparison between the observational parameters of two galaxies, NGC 147 and NGC 185. More details are available in the previous paper of this study  by  \citet{Hamedani17} and also \citet{Sohn20}. }

\begin{tabular}{llll}
 \hline
 \hline
    &     & NGC 147 & NGC 185  \\

 \hline
 \hline
Magnitude & $M_{V}$  &  $16$ & $16$     \\    
\hline
  Velocity dispersion & $ \sigma [km/s]$  &  $25$ & $25$     \\ 
\hline
Mean metallicity & $[M/H]$  &  $-1.1$ to $-1.0$ & $-1.3$ to $-1.1$     \\
\hline
Distance & $d[kpc]$  &  $675 \pm 27$ & $616 \pm 26$     \\
\hline
Distance modulus & $\mu[mag]$  &  $24.15$ & $23.95$     \\
 \hline
Distance from M31 & $d[kpc]$  &  $107$ & $160$     \\
 \hline
 \hline

\end{tabular}
\label{tab:quantities}
\end{center}
\end{table}

In this study, we aim to quantitatively test this hypothesis by estimating the mass-loss and dust injection rates in these galaxies. Dust plays a fundamental role in galaxy evolution by regulating the thermal energy balance of the ISM, catalyzing chemical reactions in molecular clouds, and contributing to star and planet formation \citep{Dwek2006}.

Pre-solar dust grains, ranging from nanometers to micrometers in size, are remnants of stardust that can be studied in laboratories. Their isotopic compositions offer insights into nucleosynthesis and stellar evolution. The most prevalent types of pre-solar grains, silicates and silicon carbides (SiC), originate from stars and are injected into the ISM via stellar winds and supernova explosions \citep{Floss16, Leitner20}.

AGB stars are a major source of interstellar dust and enriched gas, particularly during their late evolutionary stages. They experience significant mass-loss through strong stellar winds, driven by radiation pressure on dust grains and pulsation-induced shocks \citep{Hofner18}. This process ejects elements such as carbon, nitrogen, and s-process elements into the ISM, with mass-loss rates reaching up to $10^{-4} \msun yr^{-1}$ \citep{Vassiliadis93}. Consequently, AGB stars are vital contributors to the chemical enrichment of galaxies \citep{Habing13}.

Evolved stars on the Red Giant Branch (RGB) also undergo mass-loss, although at lower rates than AGB stars, playing a crucial role in stellar evolution and ISM enrichment \citep{Reimers75, Willson2000}. Observations of circumstellar dust shells around evolved stars support their contribution to the ISM \citep{Olofsson93}. Supernovae provide an additional source of dust and gas, ejecting large quantities of material into the ISM on short timescales \citep{vanLoon05a, Scalzo14}.

However, the continuous and sustained mass-loss from AGB stars makes them unique, contributing a substantial fraction of pre-solar grains found in meteorites \citep{Gail09}. Their long-term enrichment of the ISM highlights their importance in the galactic ecosystem.

This study centers on investigating the mass-loss and dust production processes associated with long-period variable stars. Previous works, such as those by \citet{Javadi13_III} and \citet{Abdollahi23}, have examined the mass-loss rates of LPV stars in the galaxies M33 and Andromeda IX, respectively, employing methodologies adapted to these stellar populations.

The primary aims of this paper are to determine the rate at which dust is produced by LPVs and transferred to the interstellar medium, and to establish correlations between the dust production rate, luminosity, and other physical properties of these stars. In this manuscript, Section \ref{sec:data} introduces the data used in our study, the models employed, and the associated codes and grid criteria. Section \ref{sec:results} explains the results. Section \ref{sec:discussion} presents the discussion, and finally, Section \ref{sec:conclusion} provides the conclusions.

%@@@@@@@@@@@@@@@@@@@@@@@@@@@@@@@@@@@@@@@@@@@@@@@@@@@@@@@@@@@@@@@@@@@@@@@@@@@@@@@@@@@@@@@@@@

\section{Data}
\label{sec:data}

\subsection{Catalogs}
\label{sec:catalogs}

Given the richness of observational data across various filter bands for the galaxies NGC 147 and NGC 185, we leveraged several well-documented catalogs to enhance our understanding of these galaxies. The utilized catalogs are outlined below. This comprehensive approach allows us to explore and analyze the mass-loss of LPV stars across a broad spectrum.
\\

\begin{figure*}
\begin{center}
\includegraphics[width=0.45\textwidth]{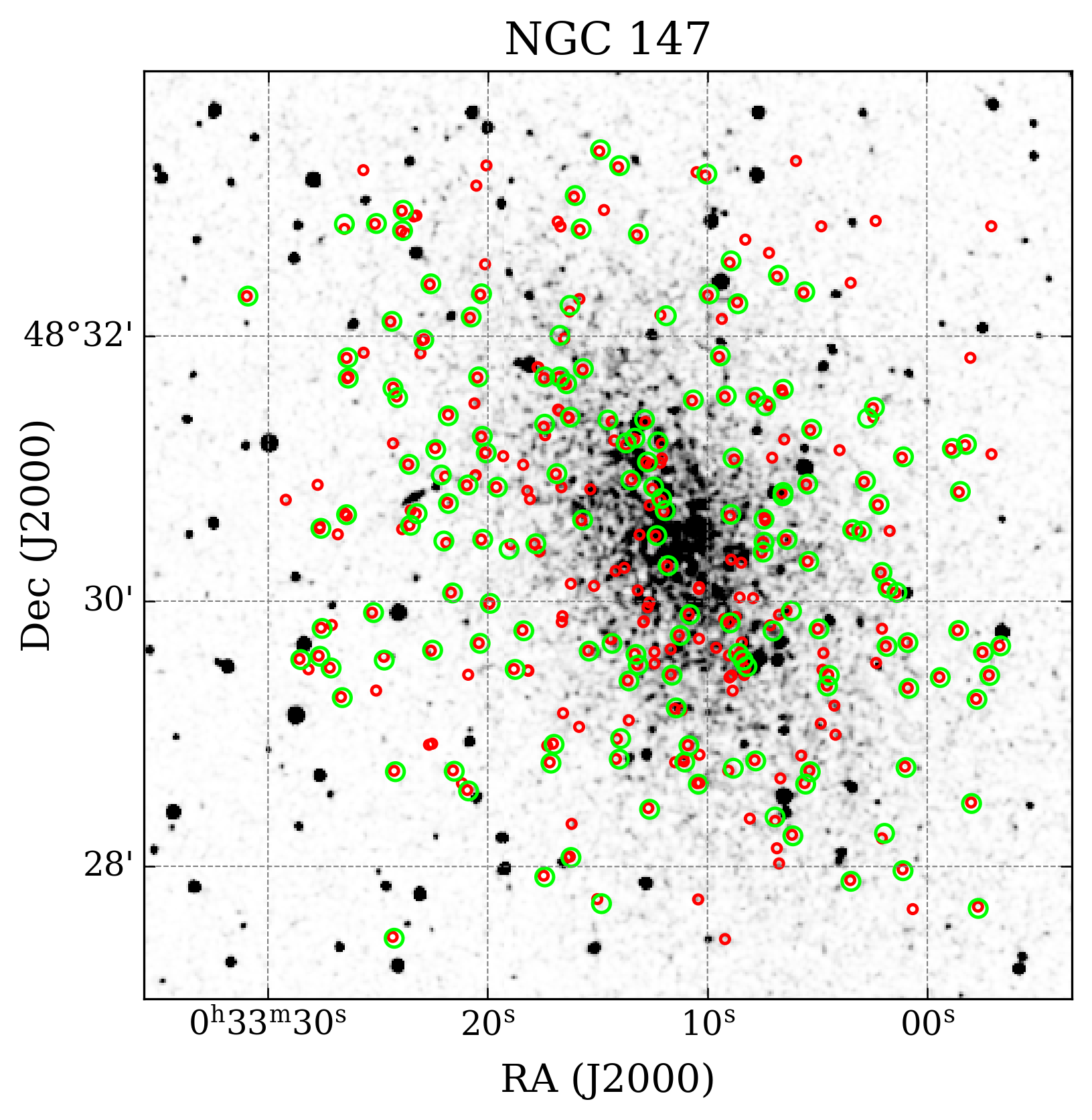}
\includegraphics[width=0.45\textwidth]{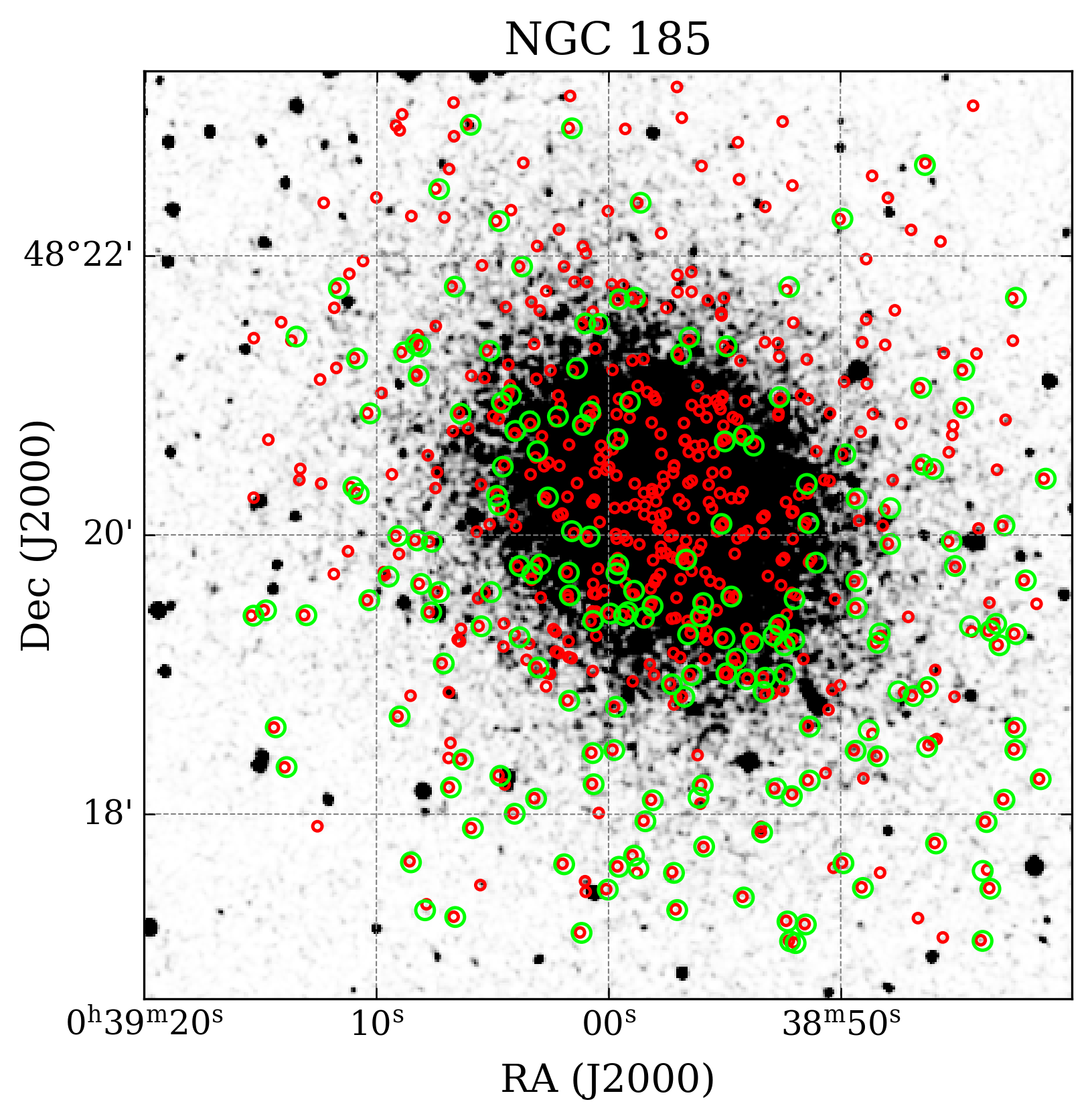}
\caption{ The distribution of long-period variable stars in the galaxies NGC 147 and NGC 185. Stars identified as LPVs from the observations of \citet{Lorenz11} are highlighted in red. Among these, stars with reported mid-infrared magnitudes for each galaxy are marked in green.}
\label{fig:DS9}
\end{center}
\end{figure*}

\begin{figure}
\begin{center}
\includegraphics[width=0.50\textwidth]{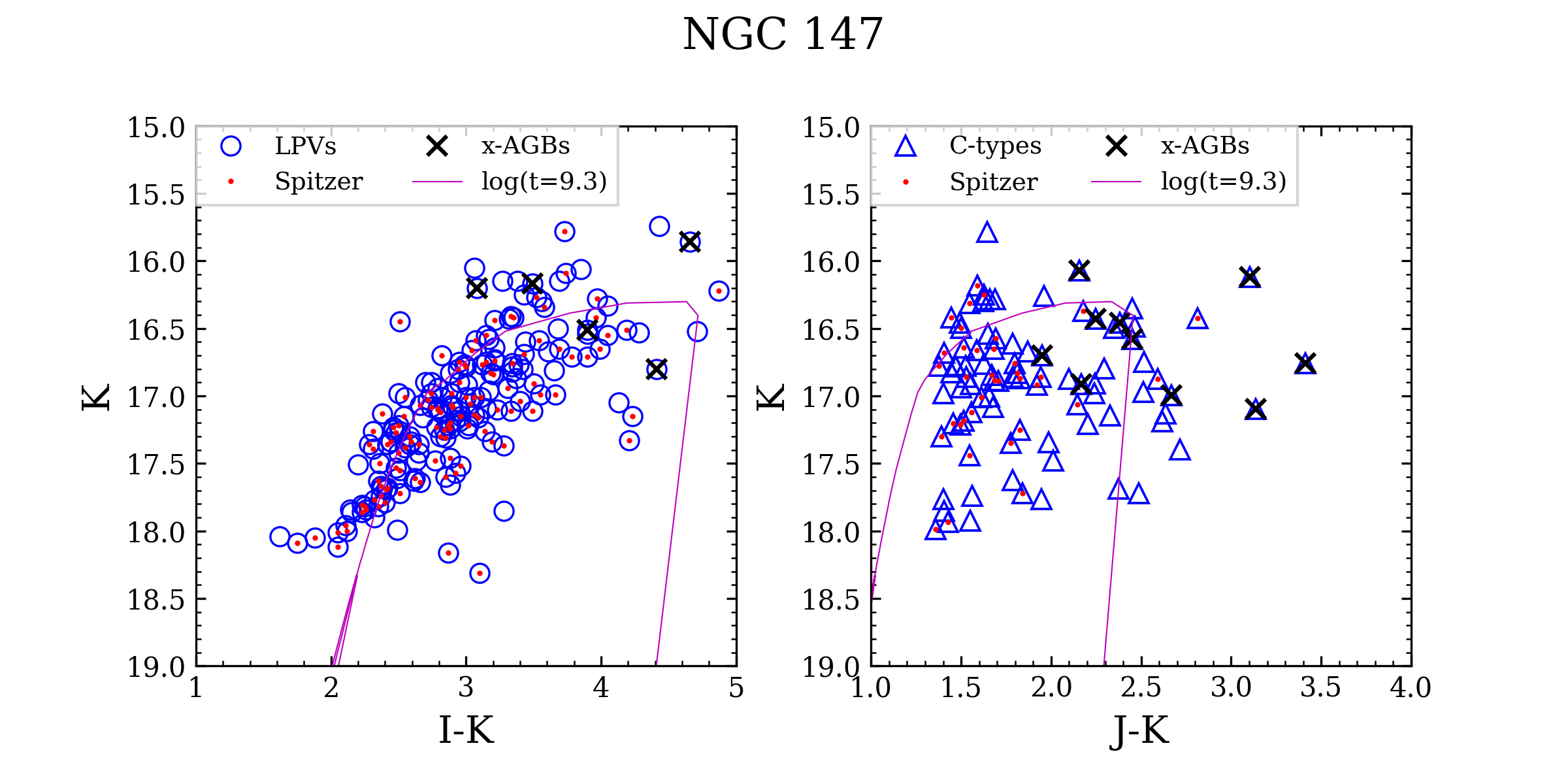}
\includegraphics[width=0.50\textwidth]{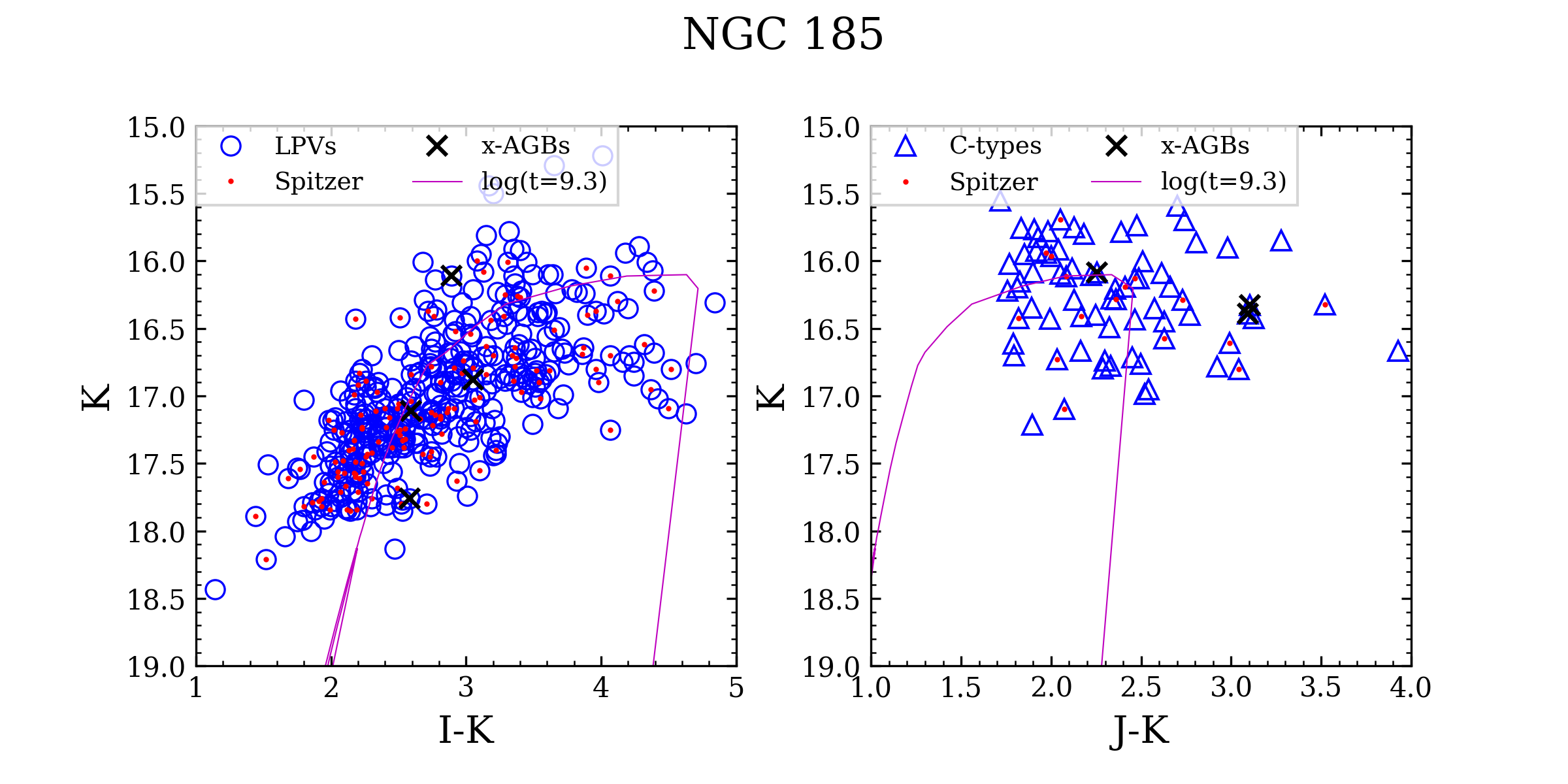}
\caption{ K-band color-magnitude diagrams showing theoretical isochrones from \citet{Marigo08} for an age of log(t) = 9.3 (equivalent to 2 Gyr) and metallicity Z=0.0015. The left-hand plots display the distribution of long-period variable stars, adapted from \citet{Lorenz11}, while the right-hand plots represent carbon-rich stars, adapted from \citet{Kang05,Sohn06}.}
\label{fig:CMD_K}
\end{center}
\end{figure}

\textbf{LPVs}: The study by \citet{Lorenz11} focuses on identifying and characterizing long-period variables in the dwarf galaxies NGC 147 and NGC 185. The primary goal of the research was to catalog LPVs, including Mira, semi-regular, and irregular variables, in these two galaxies and to analyze their light variations and period-luminosity relations (PLRs).
Using the Andalucia Faint Object Spectrograph Camera (ALFOSC) on the Nordic Optical Telescope, the team conducted time-series photometry in the i-band over approximately 2.5 years, capturing over 30 epochs. This data was complemented by single-epoch \textnormal{K$_s$}-band photometry and narrow-band photometry from the literature to distinguish between O-rich and C-rich stars. The study detected 513 LPVs in NGC 185 and 213 LPVs in NGC 147. The LPVs exhibited periods ranging from 90 to 800 days and amplitudes up to 2 magnitudes in the i-band. 
The reported photometric uncertainties ranged from 0.03 to 0.16 mag in the \textnormal{i}-band, and from 0.09 to 0.26 mag in the \textnormal{K$_s$}-band.

\textbf{Carbon stars}: \citet{Kang05} and \citet{Sohn06} cataloged the carbon stars of galaxies NGC 185 and 147. Using the CFHTIR imager and the J, H, and  K filters, they captured near-IR images with the Canada-France-Hawaii Telescope (CFHT) in June 2004. Each field had a size of $ 3.6' \times 3.6' $. The photometry of 91 carbon stars in NGC 147 is presented by \citet{Sohn06}, and the photometry of 73 carbon stars in NGC 185 is presented by \citet{Kang05}.\\

\citet{Sohn06} reported photometric uncertainties in the K filter ranging from 0.015 to 0.105 mag. The uncertainties in the J$-$K and H$-$K colors span 0.025 to 0.322 mag and 0.025 to 0.153 mag, respectively. For the NGC 185 dataset, \citet{Kang05} reported J-band uncertainties between 0.023 and 0.161 mag, J$-$K uncertainties from 0.029 to 0.163 mag, and H$-$K uncertainties from 0.023 to 0.081 mag.

\textbf{Spitzer}: DUST in Nearby Galaxies with \texttt{Spitzer} (DUSTiNGS) is an infrared survey of 50 dwarf galaxies in and around the LG designed to detect evolved stars in the dust-producing phase \citep{Boyer15a,Boyer15b,Boyer17}. 
DUSTiNGS employs the InfraRed Array Camera (IRAC), a component of the \texttt{Spitzer} Space Telescope (SST), in its post-cryogen phase. The survey had 37 dwarf spheroidals (dSph), eight dwarf irregulars (dIrr), and five dIrr/dSph transition-type galaxies. To enhance the identification of AGB stars, which exhibit variability at 3.6 \textnormal{\ensuremath{\mu m}} and 4.5 \textnormal{\ensuremath{\mu m}} wavelengths, each galaxy was examined at two distinct epochs, with a time interval of approximately six months between them. This temporal separation allows for the detection of variability, which is a hallmark of AGB stars, particularly in their late evolutionary stages. Because of the significant sample size, reliable statistics on the dust-producing, short-lived phase are possible. For example, according to this survey’s estimation of the number of dusty AGBs, NGC 147 and 185 contain roughly 124 and 99 x-AGBs, respectively. The optically obscured stars, often known as “extreme” AGBs or x-AGB stars, are typically chosen due to their red colors ([3.6]–[8] $>$ 3 mag). This color criterion helps isolate the most evolved, dust-enshrouded stars from other stellar populations \citep{Boyer15b}.  In their published catalog, \citet{Boyer15a} reported photometric uncertainties for both the 3.6 and 4.5 $\mu$m filters to range between 0.02 and 0.27 mag for sources in both galaxies.

%\textbf{Photometric uncertainties were assessed via artificial star tests and reflect the difference between input and recovered magnitudes. For NGC 147, typical uncertainties at the 75\% completeness limits are $\pm$0.2 mag at [3.6] = 18.2 mag and [4.5] = 18.3 mag, while for NGC 185 they are $\pm$0.2 mag at [3.6] = 18.6 mag and [4.5] = 18.4 mag. These completeness limits refer to the epoch 1 data; for the deeper, combined-epoch photometry, the limits are approximately 0.5 mag fainter.}

\textbf{Optical bands}: By investigating the NGC 147 and 185 galaxies through the four filters of V, i, TiO, and CN, \citet{Nowotny03} were able to determine the chemical abundance of stars. The data were obtained using the 2.56-m Nordic Optical Telescope situated in La Palma, Spain. These observations were conducted during the nights of August 30 to September 1, 2000, with the utilization of the ALFOSC instrument.
The atmosphere of AGB stars can change during evolution and under certain circumstances, such as convection or thermal pulses, and their compositions can change from O-rich to C-rich environments. C-type stars refer to AGBs with higher carbon abundance than oxygen (C/O$>$1). The other category of AGBs (C/O$<$1) is called M-type stars.

\citet{Nowotny03} defined the C- and M-type stars as follows:\\
$M :\ V-i > 1.6 \ mag,\ TiO - CN > 0.15 \ mag$\\
$C :\ V-i > 1.16 \ mag,\ TiO - CN < -0.3 \ mag$\\

They identified the type of stars using filters TiO and CN. A field of view of $ 6.5' \times 6.5' $ allowed the 
observation of 146 new C-type stars in the galaxy NGC 147 and 154 C-type stars in the NGC 185. In the first paper of this observational series, \citet{Nowotny01} reported a typical photometric uncertainty of 0.02 mag in the i band and 0.06 mag in the V$-$i color.

%\textbf{The photometric uncertainties in these measurements are significant, reflecting the variability and faintness of the AGB population. For NGC 147, the mean color and magnitude of C stars were $\left\langle V-i\right\rangle$ = 2.06 $\pm$0.44 mag and $\left\langle i\right\rangle$= 20.15 $\pm$0.42 mag while in NGC 185, these were $\left\langle V-i\right\rangle$ = 2.11 $\pm$0.40 mag and $\left\langle i\right\rangle$= 19.72 $\pm$0.46 mag.}

The two papers by \citet{Battinelli147,Battinelli185} provide a detailed examination of the carbon star populations in the galaxies NGC 147 and NGC 185. In the \citet{Battinelli185} study, the authors investigate NGC 185, identifying 145 carbon stars using the CN-TiO technique with the CFH12K wide-field camera. The mean I-magnitude of these C stars is found to be 19.99 $\pm$0.05 mag, corresponding to an absolute magnitude of $-$4.41 $\pm$0.05 mag. Photometric uncertainties for these data range from 0.010 to 0.057 mag in I, and from 0.015 to 0.109 mag in the (R–I) color index. 

%They determine the stellar surface density profile of NGC 185, which follows a power law with an exponential scale length of 2.53 $\pm$0.07 arc-minutes. Interestingly, the intermediate-age C star population has a smaller scale length, indicating it is more concentrated than the older population. They also determine a tidal radius of 22.5 $\pm$2.2 arc-minutes from red giant star counts. They find a C/M ratio of 0.17 $\pm$0.02 with no significant radial gradient, suggesting a well-mixed stellar population without significant metallicity variations across the galaxy.

%In the \citet{Battinelli147} paper, the focus shifts to NGC 147. The study identifies 288 C stars with an average I-magnitude of 20.31 $\pm$0.40 mag, corresponding to an absolute magnitude of $-$4.39 mag, slightly less luminous than other galaxies surveyed. The stellar surface density profile of NGC 147 shows a scale length of 4.1 $\pm$0.1 arc-minutes, with the tidal radius determined to be 33.9 $\pm$2.4 arc-minutes. The distribution of C stars in NGC 147 indicates that intermediate-age stars are well mixed with the older population, unlike NGC 185, which shows more concentration of intermediate-age stars. The overall C/M ratio is 0.24 $\pm$0.02, with a noticeable increase from the center to the outer regions, implying a metallicity gradient decreasing by about 0.4 dex in [Fe/H].

To compile the ultimate datasets, we utilized the LPVs of \citet{Lorenz11} catalog and the C-type stars of \citet{Kang05} and \citet{Sohn06} as the foundational references and conducted cross-matching with the remaining catalogs. The resulting composite catalog encompassed LPV stars with available data spanning at least three bands. We identified 163  LPV stars for the NGC 147 galaxy and 187 LPV stars for the NGC 185 galaxy. Figure \ref{fig:DS9} presents a comprehensive representation of stars in the final catalogs, encompassing those also observed by \citet{Boyer15a}.

Figure \ref{fig:CMD_K} illustrates the positions of LPVs in the near-IR CMDs (left panels) and compares them with carbon stars (right panels). According to \citet{Hamedani17},  the LPV stars in galaxy NGC 147 are older, with ages exceeding a few billion years, compared to those in NGC 185. The study indicate that NGC 185 contains LPV stars with ages of approximately 400 to 500 million years.

\subsection{Modeling the SEDs}
\label{sec:code}

By cross-matching the catalogs from Sec. \ref{sec:catalogs} and determining the flux of LPV stars in each galaxy, the spectral energy distributions (SEDs) of these stars can now be reconstructed. To accurately model these SEDs, it is essential to account for the chemical composition of the variable stars' atmospheres.

The quantity of carbon in the atmosphere of AGB stars increases after the third dredge-up process despite the initial abundance of oxygen \citep{Mowlavi99, Uttenthaler19}. AGB stars are typically categorized as either carbon-rich (C/O $>$ 1) or oxygen-rich (C/O $<$ 1) according to the relative quantities of oxygen and carbon in their atmospheres \citep{Nowotny03,Ferrarotti06}. Carbon stars form more easily in metal-poor environments since less carbon needs to be dredged up to achieve a C/O ratio greater than 1 \citep{Hartwig19}. Understanding the atmospheric composition of AGB stars is essential for modeling their SEDs using packages such as \texttt{DUSTY}.

\texttt{DUSTY}  is a highly versatile radiative transfer code developed by \citet{Ivezic97} designed to model the interaction of radiation with dust in various astrophysical environments. The code addresses the complex radiative transfer problem by providing solutions in a spherically symmetric geometry, allowing for the detailed study of dust-enshrouded stars, star-forming regions, and other dusty astronomical objects.
At its core, \texttt{DUSTY}  solves the radiative transfer equation, accounting for absorption, scattering, and re-emission of radiation by dust grains. This comprehensive approach enables the simulation of the SED of dust-embedded sources, facilitating the comparison with observational data. One of the critical strengths of \texttt{DUSTY}  is its ability to contain a wide range of dust grain properties, including different compositions, sizes, and optical characteristics. This flexibility is crucial for accurately modeling the diverse dust environments found in space.

\texttt{DUSTY}'s applicability extends beyond circumstellar dust shells to include broader astrophysical contexts. For instance, it can be used to model the infrared emission from star-forming regions, the extinction curves of interstellar dust, and the toroidal dust structures in active galactic nuclei \citep{Ivezic97}. The ability to generate synthetic spectra that closely match observations allows astronomers to derive physical parameters such as dust temperature, density distribution, and mass-loss rates from their models. This makes \texttt{DUSTY}  an indispensable tool for interpreting visible to sub-millimeter observations of dusty objects.

A software tool called \texttt{DESK} \citep{Goldman2020} exemplifies the utilization of the \texttt{DUSTY} kernel for extracting SEDs of variable stars. The \texttt{DESK} package is a fitting software explicitly designed for analyzing data obtained from evolved stars, encompassing both photometry and spectra. It employs radiative transfer model grids, which are generated by the \texttt{DUSTY} kernel under various initial conditions such as stellar temperature, dust grain properties, optical depth, and dust chemical composition, among other relevant factors.

However, due to the requirement for higher-resolution grids specifically for the optical depth parameter, we performed an independent parametric study to achieve the desired level of resolution. Subsequently, we compared the results obtained from our study with those generated by the \texttt{DESK} code.

We conducted an extensive parameter grid exploration, encompassing a range of models with varying free parameters, including stellar temperature, dust temperature, and optical depth. The grid covered a broad spectrum of values, with the stellar temperature spanning from 2000 to 3000K in 100K increments, the dust temperature ranging from 500 to 1500K in 100K increments, and the optical depth evolving across the interval of 0.01 to 2. Within the optical depth range of 0.01 to 0.5, increments of 0.01 were employed, while within the range of 0.5 to 2, each increment was set at 0.05. 
The chemical composition of the grid was defined based on the type of dust species used to achieve the best fits. For carbon stars, we assumed a grain mixture comprising 85\% amorphous carbon \citep{Hanner88} and 15\% silicon carbide \citep{Pegourie88}. For oxygen-rich AGB stars, we adopted astronomical silicates \citep{Draine84}.

This comprehensive grid was executed separately for oxygen-rich and carbon stars to account for variations in stellar compositions. Consequently, within the ultimate sample of stars, a total of 12,000 oxygen models and an equivalent number of carbon models were considered. This approach allowed us to capture the differences between the two types of stars, ensuring that our models could be effectively compared with those produced by the \texttt{DESK} code.

The process of selecting the most appropriate model for each star involved employing the method of least chi-square. This statistical method enabled us to identify the model whose SED best aligned with the observational data for each individual star. By minimizing the chi-square value, we ensured that our models provided the closest possible fit to the observed SEDs, thereby enhancing the reliability of our comparison with \texttt{DESK}.

%\textbf{To ensure a robust and unbiased comparison between our models and the observational data, all photometric measurements—regardless of filter similarity—were treated as independent flux points in the SED fitting process. Filters with comparable effective wavelengths and bandwidths, such as the i- and I-band or the \textnormal{K$'$}- and \textnormal{K$_s$}-band, were not averaged or combined. Instead, magnitudes were converted to physical fluxes using filter-specific zero points and effective wavelengths. For example, the i-band (e.g., from the Nordic Optical Telescope) has a zero point of 2435.41 Jy and an effective wavelength of 792.7 nm, whereas the I-band (e.g., from CFHT) has a zero point of 2407.05 Jy and an effective wavelength of 815.4 nm. Similarly, the \textnormal{K$'$}-band (zero point: 686.6 Jy  $\lambda_{\text{eff}} = 2105.4$ nm) and \textnormal{K$_s$}-band (zero point: 667.0 Jy, $\lambda_{\text{eff}} = 2138.2$ nm) were treated separately.}

%\textbf{Each flux measurement was included independently in the $\chi^2$ minimization process, with all measurements contributing equally to the SED fitting. This method ensures that differences in calibration, variability, or unique spectral features for each filter are properly accounted for. As a result, the best-fit SED model may align more closely with the flux from one filter (e.g., i-band) compared to a similar filter (e.g., I-band).}

Given that the stars in our catalog have reached the final stage of their evolution, their luminosity can be correlated with their initial mass \citep{Kippenhahn90}. In instances where the literature lacked prior information on the chemical composition of a star, we utilized the star’s initial mass to ascertain this characteristic \citep{Renzini81,Groenewegen93,Girardi07,Leisenring08}. The mass range for stars to become carbon-rich AGB stars varies with metallicity. At solar metallicity ([M/H]=0), stars with initial masses between approximately $2\msun$ and $4.5\msun$ can become carbon stars through the third dredge-up process \citep{Karakas14_II}. As metallicity decreases, the mass range for carbon star formation shifts to lower masses \citep{Marini21}. 

Considering the metallicity of the two target galaxies in this study, as detailed in Table \ref{tab:quantities}, the initial mass range for carbon stars was determined to lie between 1.1$\msun$ and 4$\msun$, with stars outside this range categorized as oxygen-rich stars \citep{Herwig05,Javadi11_II, Marini21, Saremi21}.

%Figure \ref{fig:2x2} presents a representative plot of the outputs obtained from our parameter grid. Employing the minimum chi-square method, we selected the SED that most closely aligns with the observational data. The parameters associated with this optimal SED are also reported (Table \ref{tab:selected_info} and Appendix \ref{sec:apndix}).

\begin{figure}
\begin{center}
\includegraphics[width=0.50\textwidth]{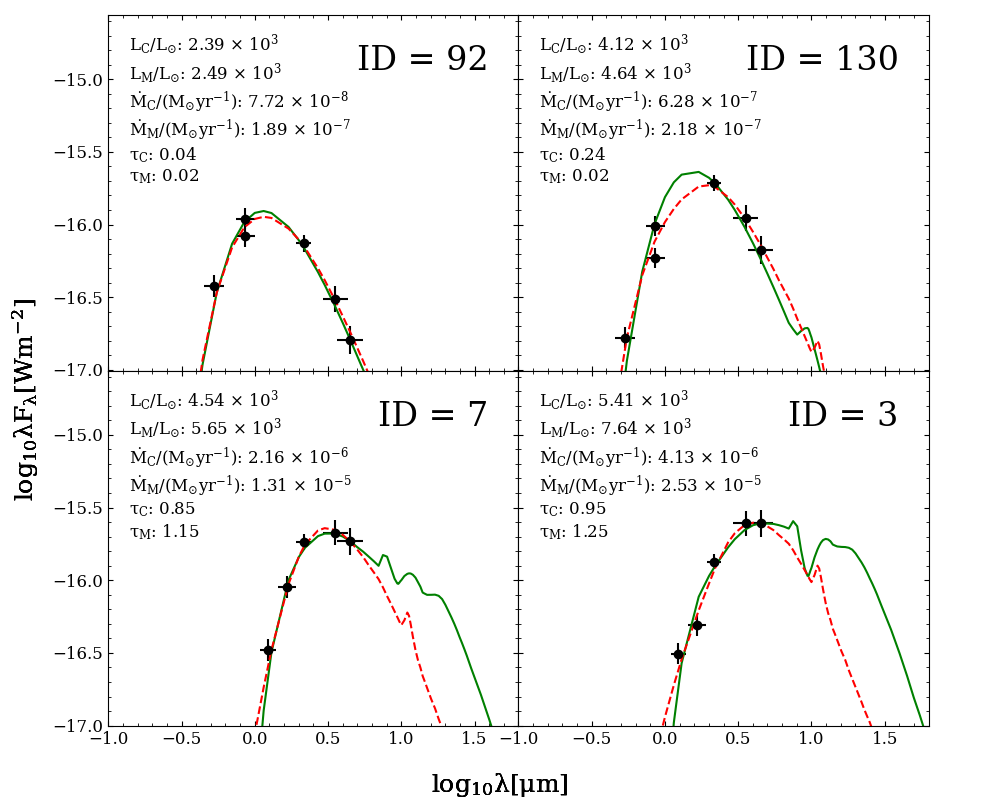}
\caption{ Grid results for four stars within the NGC 147 galaxy. The solid green line depicts the oxygen models, while the red dashed line illustrates the carbon models. Error bars represent the photometric uncertainties for each band, as provided in the original catalogs.}
\label{fig:2x2}
\end{center}
\end{figure}

Building on the approach described above, we constructed a grid of models and applied a minimum chi-square fitting procedure to identify the SED that best matches the observed photometry. Figure~\ref{fig:2x2} presents a representative example of the model fits, and the parameters corresponding to the best-fit SEDs are summarized in Table~\ref{tab:selected_info} and detailed in Appendix~\ref{sec:apndix}.

\begin{table*}
\begin{center}
 \caption{  Summary of LPV properties by chemistry. This table summarizes the calculated optical depth, luminosity, and mass-loss rates for C-type and M-type LPV stars shown in Figure \ref{fig:2x2}. A more comprehensive table, including these stars and additional stars from the NGC 147 and NGC 185 catalogs, is provided in Appendix \ref{sec:apndix}.}
%\resizebox{0.49\textwidth}{!}{
\begin{tabular}{ccccccc}
\hline
\hline
ID & \textnormal{\ensuremath{\tau}\textsubscript{C}} & \textnormal{\ensuremath{\tau}\textsubscript{M}}& \textnormal{L\textsubscript{C}} & \textnormal{L\textsubscript{M}} & \textnormal{\ensuremath{\dot{M}}\textsubscript{C}} &  \textnormal{\ensuremath{\dot{M}}\textsubscript{M}}  \\ 
 & & &$[\lsun]$ &$[\lsun]$ & $[\msun yr^{-1}$] & [$\msun yr^{-1}$] \\
\hline
\hline
 3   & 0.95 $\pm$ 0.05 & 1.25 $\pm$ 0.05 & $(5.41 \pm 0.23) \times 10^3$ & $(7.64 \pm 0.31) \times 10^3$ & $(4.13 \pm 1.29) \times 10^{-6}$ & $(2.18 \pm0.90) \times 10^{-7}$ \\
 \hline
 7   & 0.85 $\pm$ 0.05 & 1.15 $\pm$ 0.05 & $(4.54 \pm 0.19) \times 10^3$ & $(5.65\pm 0.23) \times 10^3$ & $(2.16 \pm 0.67) \times 10^{-6}$ & $(1.31 \pm 0.54) \times 10^{-5}$ \\
 \hline
 92  & $0.04 ^{+0.05}_{-0.04}$ & $0.02^{+0.05}_{-0.02}$ & $(2.39 \pm 0.10) \times 10^3$ & $(2.49 \pm 0.10) \times 10^3$ & $(7.72 \pm 2.40) \times 10^{-8}$ & $(1.89 \pm 0.78) \times 10^{-7}$ \\
 \hline
 130 & 0.24 $\pm$ 0.05 & $0.02^{+0.05}_{-0.02}$ &$(4.12 \pm 0.17) \times 10^3$  &$(4.64 \pm 0.19) \times 10^3$ &  $(6.28 \pm 1.96) \times 10^{-7}$ &  $(2.18 \pm 0.90) \times 10^{-7}$ \\
\hline
\hline
\end{tabular}
%}
\label{tab:selected_info}
\end{center}
\end{table*}

The uncertainties from the \texttt{DUSTY} fitting procedure arise from two primary sources: systematic effects due to model assumptions, and statistical uncertainties arising from photometric errors. A systematic uncertainty of 30\% on the mass-loss rates was adopted, following the modeling assumptions and limitations discussed in \texttt{DUSTY} \citep{Ivezic97}. This value reflects the inherent insensitivity of the SEDs fitting to certain physical parameters, such as the gravitational correction factor.

To estimate statistical uncertainties, we applied random perturbations to the observed magnitudes of representative stars, based on their photometric errors and assuming Gaussian distributions. For the star with ID 43 in NGC 147, 50 such realizations were generated and individually fitted with \texttt{DUSTY}, for both carbon-rich and oxygen-rich models. This star was selected because it owns magnitudes in all available filter bands. The standard deviations of the resulting model parameters were adopted as measures of the statistical error. For the carbon-rich model, the statistical uncertainty in the mass-loss rate was $\pm 2.50 \times 10^{-8}  M_\odot  \text{yr}^{-1}$ (8.45\% relative error) for a reference mass-loss rate of $\dot{M} = 2.96 \times 10^{-7}  M_\odot  \text{yr}^{-1}$. The luminosity uncertainty was $\pm 1.30 \times 10^{2}  L_\odot$ (4.25\% relative error) for a reference luminosity of $L = 3.07 \times 10^{3}  L_\odot$. For the oxygen-rich model, the statistical uncertainties were 28.26\% for the mass-loss rate and 4.05\% for the luminosity, derived using the same methodology.

The total uncertainty in the mass-loss rate was calculated by merging the statistical and systematic uncertainties using the standard error combination method:

\begin{equation}
\sigma_{\text{tot}} = \sqrt{ \sigma_{\text{stat}}^2 + \left(0.3 \times \dot{M}_{\text{tot}}\right)^2 }
\label{eq:tot_err}
\end{equation}

For the carbon-rich model of the reference star, this yielded a total mass-loss uncertainty of $\pm 9.21 \times 10^{-8}  M_\odot  \text{yr}^{-1}$ ($31.1\%$). For oxygen-rich stars, with a statistical uncertainty of $28.26\%$, the total uncertainty becomes $41.20\%$.  For NGC 185, where only two stars (IDs 18 and 49) had complete available filter observations, the same statistical uncertainties were adopted due to the comparable observational instruments and distances of NGC 147 and NGC 185. Thus, the uncertainties for carbon-rich and oxygen-rich models in NGC 185 mirror those of NGC 147.

As it should be noted, accurate construction of SEDs for $\chi^2$ minimization required careful handling of multi-band photometric data, accounting for systematic differences between similar filters and stellar variability. This approach was essential to ensure that the observational data accurately reflected the physical properties of stars, particularly for filters with overlapping wavelength ranges.

The i/I and K/K$_s$ filter pairs have nearly identical effective wavelengths and bandwidths but show systematic magnitude differences of $\sim$0.5–1 mag for i/I and $\sim$0.15 mag for K/K$_s$ due to filter transformation effects \citep{Jordi06,Leggett06}. The i and I filters differ in their response curves, and with R–I colors of 1–2 mag in our data, offsets up to $\sim$1 mag are expected \citep{Jordi06}.  In some instances, larger discrepancies exceeding typical uncertainties highlight the impacts of variability, such as star ID 57 in NGC 147, which exhibits a 1.29 mag difference between i (20.95 mag) and I (19.66 mag).

In SED fitting, we explicitly avoided averaging magnitudes of similar filters (e.g.,i$/$I or K$/$K$_s$) to prevent biases in model comparisons. Instead, each photometric measurement was treated as an independent flux point, converted from magnitudes using filter-specific zero points and effective wavelengths, such as i-band ($2435.41$ Jy, $792.7$ nm), I-band ($2407.05$ Jy, $815.4$ nm), K-band ($686.6$ Jy, $2105.4$ nm), and K$_s$-band ($667.0$ Jy, $2138.2$ nm). By adopting these unique filter characteristics, each flux point contributed equally to the $\chi^2$ minimization, ensuring that calibration differences, stellar variability, and distinct spectral features were accurately captured in the best-fit SED models.

For all stars in our final catalogs, including those like star ID 57 with significant magnitude differences between i and I filters, SED fits using both filters yield mass-loss rates of $1.55 \times 10^{-7}  M_\odot  \text{yr}^{-1}$ for carbon-rich models and $8.66 \times 10^{-8}  M_\odot  \text{yr}^{-1}$ for oxygen-rich models. Fits using only the $i$-band ($20.95$ mag) or $I$-band ($19.66$ mag) result in deviations of up to $20.1\%$ for carbon-rich models and $36.7\%$ for oxygen-rich models, which are within the SED calculations uncertainties of $31.1\%$ and $41.2\%$, respectively.

\begin{figure}
\begin{center}
\includegraphics[width=0.50\textwidth]{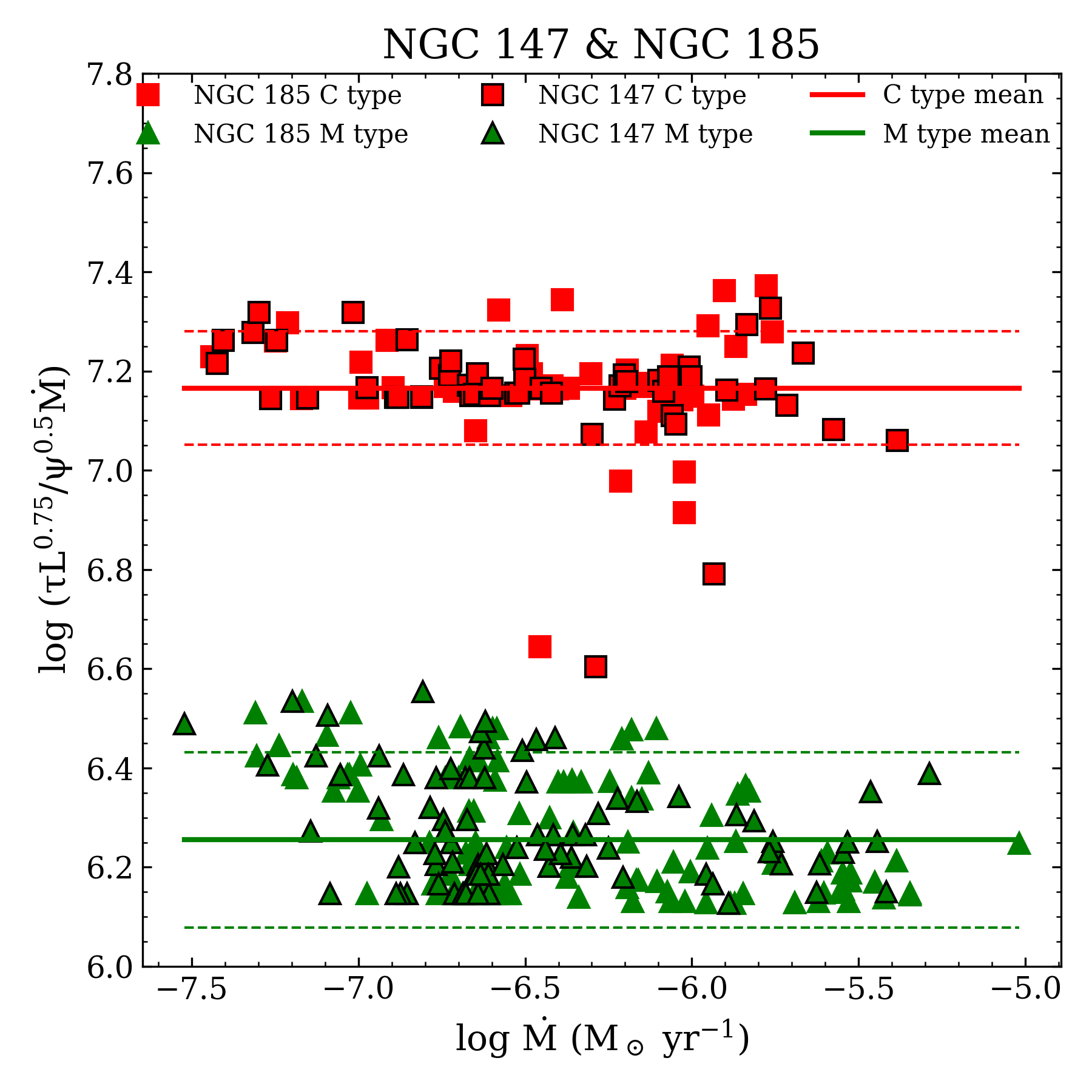}
\caption{ Relationship between the combination of optical depth, stellar luminosity, gas-to-dust mass ratio, and mass-loss rate (as defined by Equation~\ref{eq:mdot_const}) and the derived mass-loss rate. Red markers represent carbon-rich stars, while green markers indicate oxygen-rich stars. The dashed lines denote the 1$\sigma$ deviation from the mean value.}
\label{fig:C_cnst}
\end{center}
\end{figure}

Another point to consider and emphasize is that \texttt{DUSTY}'s outputs are computed for stars with a luminosity equivalent to $10^{4} \lsun $, and it calculates the mass-loss proportionate to this luminosity. As shown in Figure \ref{fig:C_cnst},  \citet{Javadi13_III} demonstrated that within a slight dispersion range, the value of the following equation remains relatively constant. 

\begin{equation}
    C= \frac{\left(\tau L^{3/4}\right)}{\left(\psi^{1/2}\dot{M}\right)} 
    \label{eq:mdot_const}
\end{equation}

In this context, C is an environment-dependent constant \citep{Javadi13_III}, $\tau$ represents the optical depth, $L$ denotes the luminosity, $\psi = 160$  \citep{Marleau10} denotes the gas-to-dust mass ratio of the galaxy, and $ \dot{M} $ signifies the stellar mass-loss. By applying this relation and the actual luminosity of each star, we scaled the mass-loss rates accordingly. As we will describe in Section~\ref{sec:ML_AGBs}, this formulation is also used to estimate mass-loss rates for stars that do not have observations in the minimum number of filter bands, needed to create reliable SEDs.

Given the similar distances of NGC 147 and NGC 185, both galaxies were analyzed together. As shown in Figure~\ref{fig:C_cnst}, their data significantly overlap, justifying a combined analysis. The standard deviation of the constant C is $0.11$ dex for carbon-rich stars and $0.17$ dex for oxygen-rich stars, corresponding to mass-loss rate uncertainties of approximately $29\%$ for carbon stars and $48\%$ for oxygen stars.

%@@@@@@@@@@@@@@@@@@@@@@@@@@@@@@@@@@@@@@@@@@@@@@@@@@@@@@@@@@@@@@@@@@@@@@@@@@@@@@@@@@@@@@@@@@

\section{Results}
\label{sec:results}

Utilizing the photometric catalogs containing information on LPVs, as detailed in Section \ref{sec:catalogs}, and employing the methodology outlined extensively in Section \ref{sec:code}, we calculate the luminosity, mass-loss, and other characteristics of these evolved stars within the environments of NGC 147 and NGC 185.

\subsection{Luminosity}

%The primary challenge in investigating the relationship between mass-loss rate and luminosity is the difficulty in finding a direct dependence of the mass-loss rate on luminosity in the observational methods. This challenge is because of the simultaneous variations of other parameters, including mass, effective temperature, and pulsation characteristics, all of which undergo changes during AGB evolutions \citep{Prager22}. It is important to note that, despite the overall positive correlation between mass-loss rate and luminosity, the data shows a significant variations.

Higher stellar luminosity is theoretically predicted to increase radiation pressure on circumstellar dust grains, thereby driving stronger stellar winds and elevating mass-loss rates in AGB stars \citep{Habing13,Hofner18}. This positive correlation between luminosity and mass-loss rate is well-supported by theoretical models, which propose that mass loss results from the interplay between radiation pressure on dust and the star's gravitational pull \citep{vanLoon99, Srinivasan09}. As a star becomes more luminous, its outward photon flux increases, effectively boosting radiation pressure and facilitating the levitation and ejection of the outer atmospheric layers. This mechanism underlies the expected rise in mass-loss rate with increasing luminosity.

 However, observational data reveal considerable dispersion in the mass-loss–luminosity relation.  A key challenge in quantifying this relationship lies in the difficulty of isolating a direct dependence of mass-loss rate on luminosity, due to the concurrent variations of several other stellar parameters (such as mass, effective temperature, and pulsation properties) that evolve simultaneously during the AGB phase \citep{Prager22}. For instance, \citet{Goldman17} developed an empirical mass-loss prescription that incorporates luminosity, pulsation period, gas-to-dust mass ratio, and the metallicity . Their analysis, particularly in equations 3 to 6, demonstrates that while luminosity shows the strongest correlation with mass-loss rate, other parameters can also modulate the efficiency of mass-loss. Moreover, they found that the luminosity dependence of mass-loss is consistent across samples from the Large Magellanic Cloud (LMC), the Galactic Centre, and the Galactic bulge, though there is some indication that this dependence may weaken slightly at higher metallicities.

\begin{figure*}
\begin{center}
\includegraphics[width=0.45\textwidth]{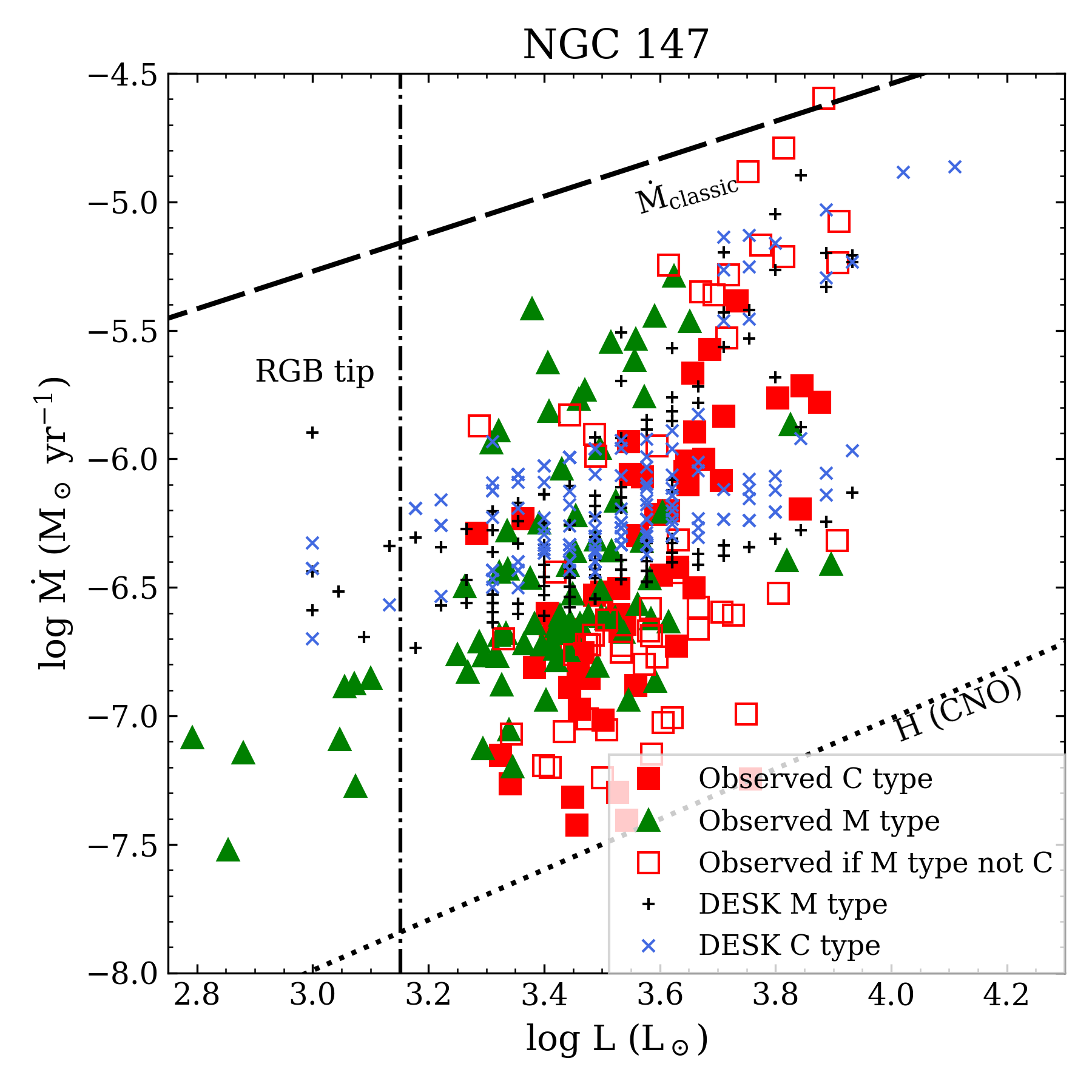}
\includegraphics[width=0.45\textwidth]{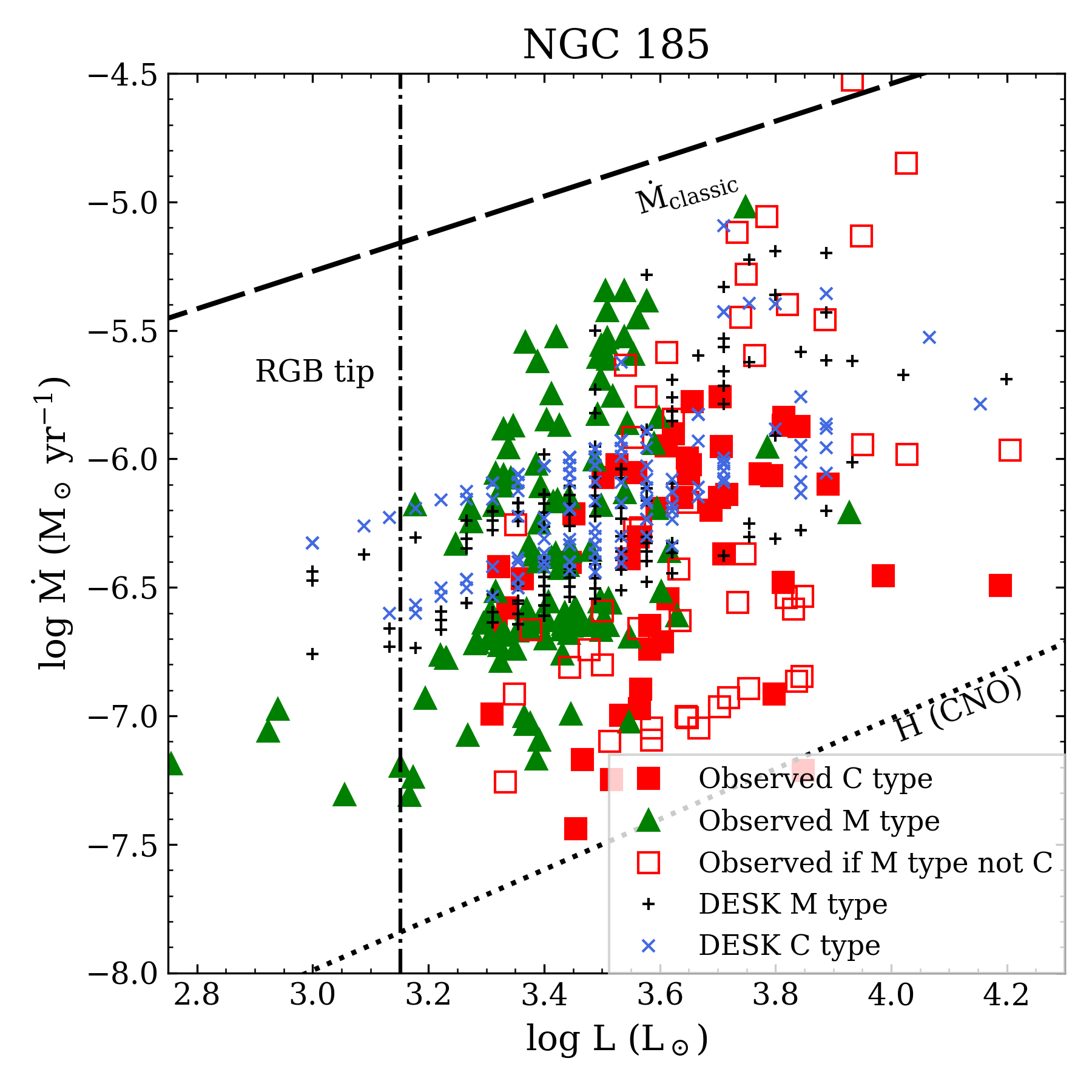}
\caption{ Mass-loss rate as a function of luminosity for LPV stars. Low-mass M-type AGB stars (green triangles) and intermediate-mass AGB carbon stars (red squares) are represented in the mass-loss rate versus luminosity model using \texttt{DUSTY}. The \texttt{DESK}-code outcomes for these stars are displayed as black pluses and blue crosses, respectively. If the carbon stars are assumed to be oxygen-rich instead, the results are displayed in the open red squares. The diagonal dotted line indicates the mass-consumption rates by CNO cycle burning on the AGB; the vertical dash-dotted line indicates the tip luminosity of the RGB; the diagonal dashed line indicates the classical limit of the mass-loss rate in dust-driven winds due to single scattering \citep{vanLoon99}.} 
\label{fig:L_MDOT}
\end{center}
\end{figure*}

Figure \ref{fig:L_MDOT} illustrates the mass-loss rates of C-rich (represented by red squares) and O-rich LPVs (indicated by green triangles). Using Table 1 from \citet{Hamedani17}'s study, the maximum initial mass for long period variable stars in our catalog is $2.48\msun$ for NGC 147 and $3.03\msun$ for NGC 185. Notably, our final catalogs do not include stars exceeding $4\msun$. Within this mass range, carbon stars are more massive than oxygen-rich stars, leading to higher luminosities. Additionally, carbon stars are expected to experience greater mass-loss than oxygen-rich stars due to their higher optical depths, as demonstrated by formula 8 in \citet{Hofner18}. This behavior is effectively illustrated in Figure 27 of \citet{Abdollahi23}.

%Figure~\ref{fig:L_MDOT} illustrates the mass-loss rates of carbon-rich (C-rich; red squares) and oxygen-rich (O-rich; green triangles) LPVs in our sample. According to Table 1 of \citet{Hamedani17}, the maximum initial mass for LPVs in our catalog is $2.48,M_\odot$ for NGC 147 and $3.03,M_\odot$ for NGC 185, with no stars exceeding $4,M_\odot$ included in the final dataset. Within this mass range, carbon stars are generally more massive—and hence more luminous—than oxygen-rich stars. As a result, C-rich LPVs are expected to experience higher mass-loss rates, further amplified by their typically higher optical depths, as predicted by formula 8 in \citet{Hofner18}. This trend is clearly illustrated in Figure 27 of \citet{Abdollahi23}.

However, Figure \ref{fig:L_MDOT} reveals a mixture of carbon and oxygen stars in both galaxies. This mixing phenomenon can be attributed to the fact that carbon stars have not yet reached the end of their AGB evolutionary phase. As time progresses, these stars are expected to transition towards higher luminosities, subsequently leading to more significant mass-losses.

When applying silicate-based models to carbon stars, it is typically observed that more significant mass-loss rates are obtained (open red squares). This phenomenon can be elucidated by the fact that silicate's optical depth is lower than carbon. Consequently, a higher column density of silicates is needed to produce an equivalent level of opacity. The direct correlation between mass-loss and optical depth will be discussed in more detail in the Section \ref{Sec:tau_color}.

Figure \ref{fig:L_MDOT} portrays the upper boundary of the classical limit of mass-loss rates \citep{vanLoon99}. As shown in this figure, the \texttt{DESK} code's results have less dispersion and occupy a smaller area of the luminosity-mass-loss space. This results from the code's lower resolution compared to our grid, which has produced identical outcomes for various stars. With regard to the oxygen stars in the NGC 147 galaxy, for instance, \texttt{DESK} has calculated the same luminosity for stars with IDs of 34, 58, 60, 61, 63, 86, 92, 93, 96, 97, 103, 110, 128, 137, 144, and 151. Of these, the stars with ID 34 and 63, 86, 103, and 151 all had mass-losses and overlapped on the luminosity–mass-loss plane.

For galaxy NGC 147, the luminosity range spans from $(6.1 \pm0.25) \times 10^{2} \lsun$ to $(7.8 \pm 0.32) \times 10^{3} \lsun$, while for galaxy NGC 185, it extends from $(5.7 \pm 0.23) \times 10^{2} \lsun$ to $(1.6 \pm 0.07) \times 10^{4} \lsun$. The range of mass-loss in these two galaxies falls within the intervals of $(1.4 \pm 0.58) \times 10^{-8} \msun yr^{-1}$ to $(5.1 \pm 1.59) \times 10^{-6} \msun yr^{-1}$, and $(1.1 \pm 0.34) \times 10^{-8} \msun yr^{-1}$ to $(9.5 \pm 3.91) \times 10^{-6} \msun yr^{-1}$, respectively. More details are presented in Table \ref{tab:Max-min}. Following the theoretical criterion of \citet{Neilson08,Neilson09}, stars with \(\log_{10} (\dot{M} / \msun) < -7.5\) were excluded, as such low mass-loss rates are more typical of Cepheid variables than of LPVs. Since star formation in NGC 185 has persisted closer to the present \citep{Hamedani17}, this galaxy hosts more massive stars, accounting for the higher stellar mass-loss rates observed compared to NGC 147.

Carbon stars within the Milky Way can achieve mass-loss rates of less than $ 10^{-4} \msun yr^{-1}$ \citep{Whitelock06}, whereas those in the Large Magellanic Cloud and M33 attain rates around $ 10^{-5} \msun yr^{-1}$ \citep{Gullieuszik12, Javadi13_III}, which is comparable to the carbon stars in both of our targets. M-type AGB stars in the solar neighborhood likewise achieve similar mass-loss rates, a few $ 10^{-5} \msun yr^{-1}$ \citep{Jura89}, which aligns with our results, too.

\begin{table*}
\begin{center}
 \caption{Minimum and maximum values calculated in different models of LPV stars. }
\begin{tabular}{lllllllll}
\hline
\hline
\multirow{3}{*}{}                 & \multicolumn{4}{c|}{NGC 147}                                                                                          & \multicolumn{4}{c}{NGC 185}                                                                                          \\ \cline{2-9} 
                                  & \multicolumn{2}{c|}{GRID}                                 & \multicolumn{2}{c|}{DESK}                                 & \multicolumn{2}{c|}{GRID}                                 & \multicolumn{2}{c}{DESK}                                 \\ \cline{2-9} 
                                  & \multicolumn{1}{c|}{M-type} & \multicolumn{1}{c|}{C-type} & \multicolumn{1}{c|}{M-type} & \multicolumn{1}{c|}{C-type} & \multicolumn{1}{c|}{M-type} & \multicolumn{1}{c|}{C-type} & \multicolumn{1}{c|}{M-type} & \multicolumn{1}{c}{C-type} \\ \hline
\hline                                  
\multicolumn{1}{l}{Luminosity $[10^3 \lsun]$}   &     0.6 - 7.8 & 2.5 - 6.3  & 1.0 - 8.5      &  1.0 - 12.9  &   0.5 - 8.4  & 1.4 - 15.4 &    1 - 15.8   & 1.0 - 14.2     \\
\hline
\multicolumn{1}{l}{Mass-loss $[10^{-7} \msun yr^{-1}] $}     &  0.3 - 51.6 &  0.1 - 26.6  &   1.8 - 126.7 & 2.0 - 136.6  &  0.5 - 95.8  &   0.1 - 17.4  & 1.7 - 64.5                             &   2.5 - 80.6     \\
\hline   
\multicolumn{1}{l}{Optical depth} &   0.01 - 0.36  &  0.01 - 0.95  & 0.1 - 2.52   &  0.1 - 1.32  &  0.01 - 0.39  &  0.01 - 0.90  &    0.1 - 1.61    &   0.1 - 0.71     \\
\hline                                  
\hline                                                
\end{tabular}
\label{tab:Max-min}
\end{center}
\end{table*}

\subsection{Optical depth}
\label{Sec:tau_color}

The luminosity of a star provides insights into its mass, radius, and evolutionary state. For AGB stars, however, the dense circumstellar envelope can obscure the observed luminosity due to self-extinction, where dust absorbs and re-emits radiation at longer wavelengths. This effect is more pronounced at shorter wavelengths and depends on the envelope's density, gas-to-dust ratio, and optical depth \citep{Whitelock06}.

Mass-loss rate in AGB stars is driven by the interaction between dust opacity and the stellar radiation field. Dust grains absorb radiation and are accelerated outward, with the efficiency of this process increasing with optical depth \citep{Bladh12}. This radiation pressure, combined with pulsation-driven shock waves, contributes to the overall mass-loss rate.

\begin{figure*}
\begin{center}
\includegraphics[width=0.45\textwidth]{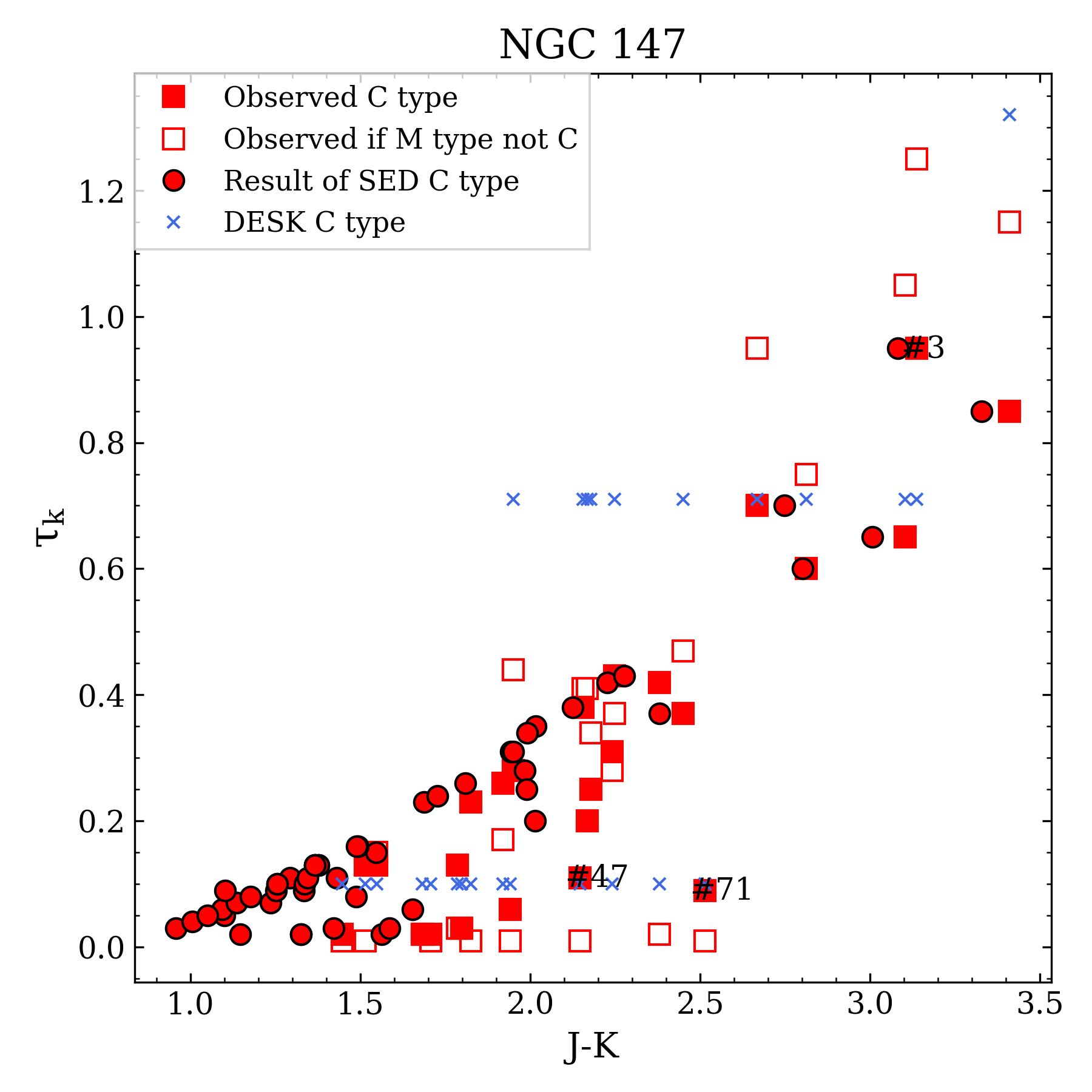}
\includegraphics[width=0.45\textwidth]{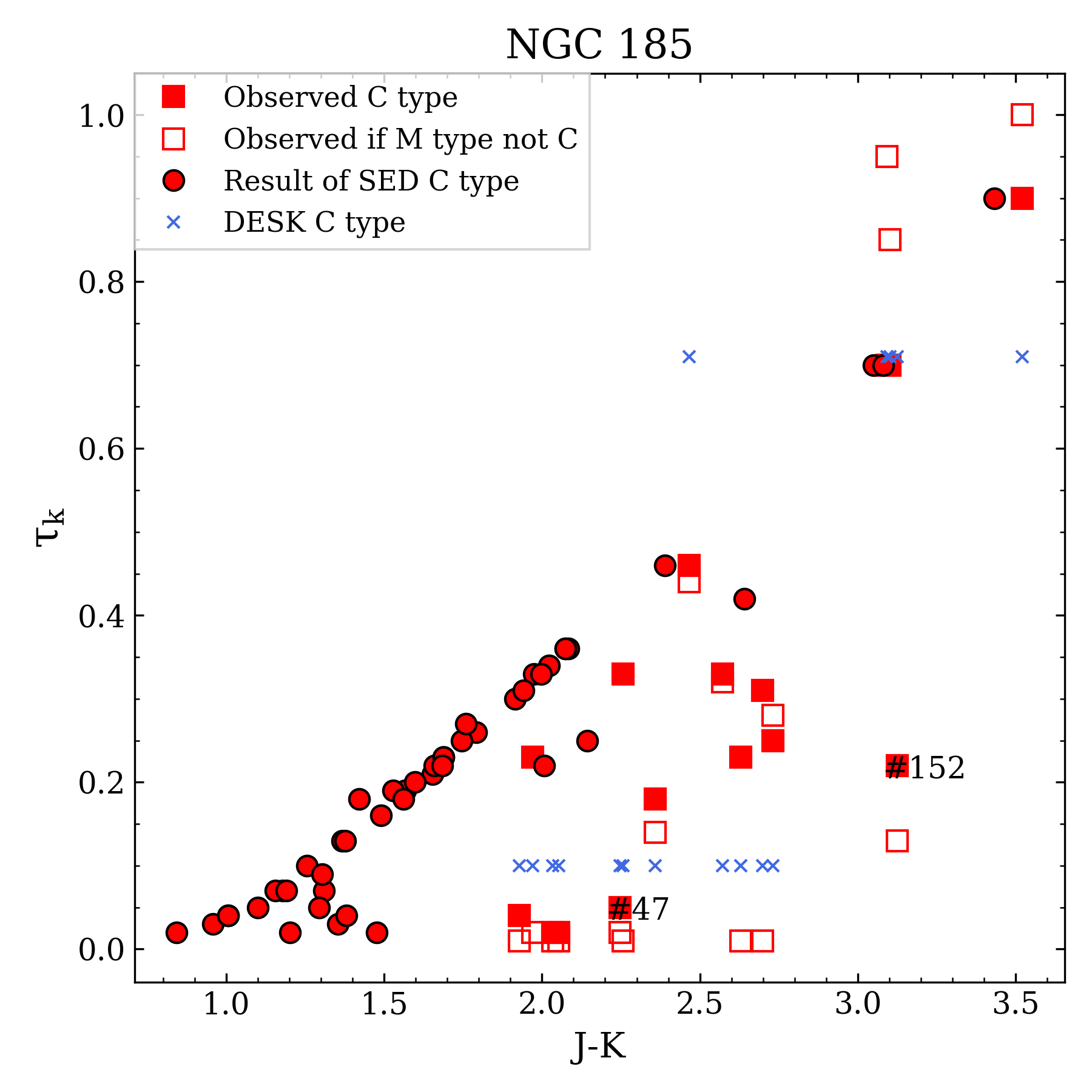}
\includegraphics[width=0.45\textwidth]{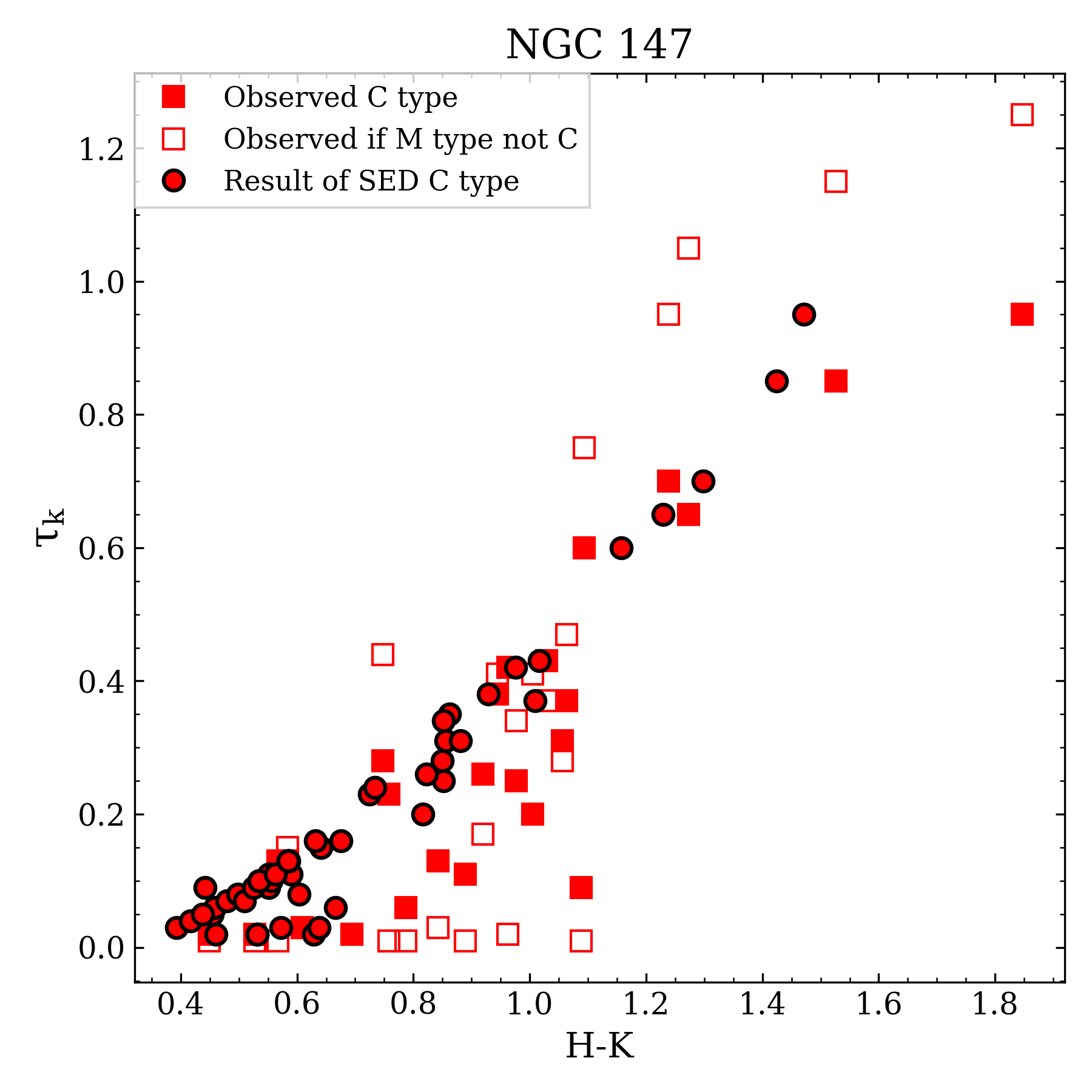}
\includegraphics[width=0.45\textwidth]{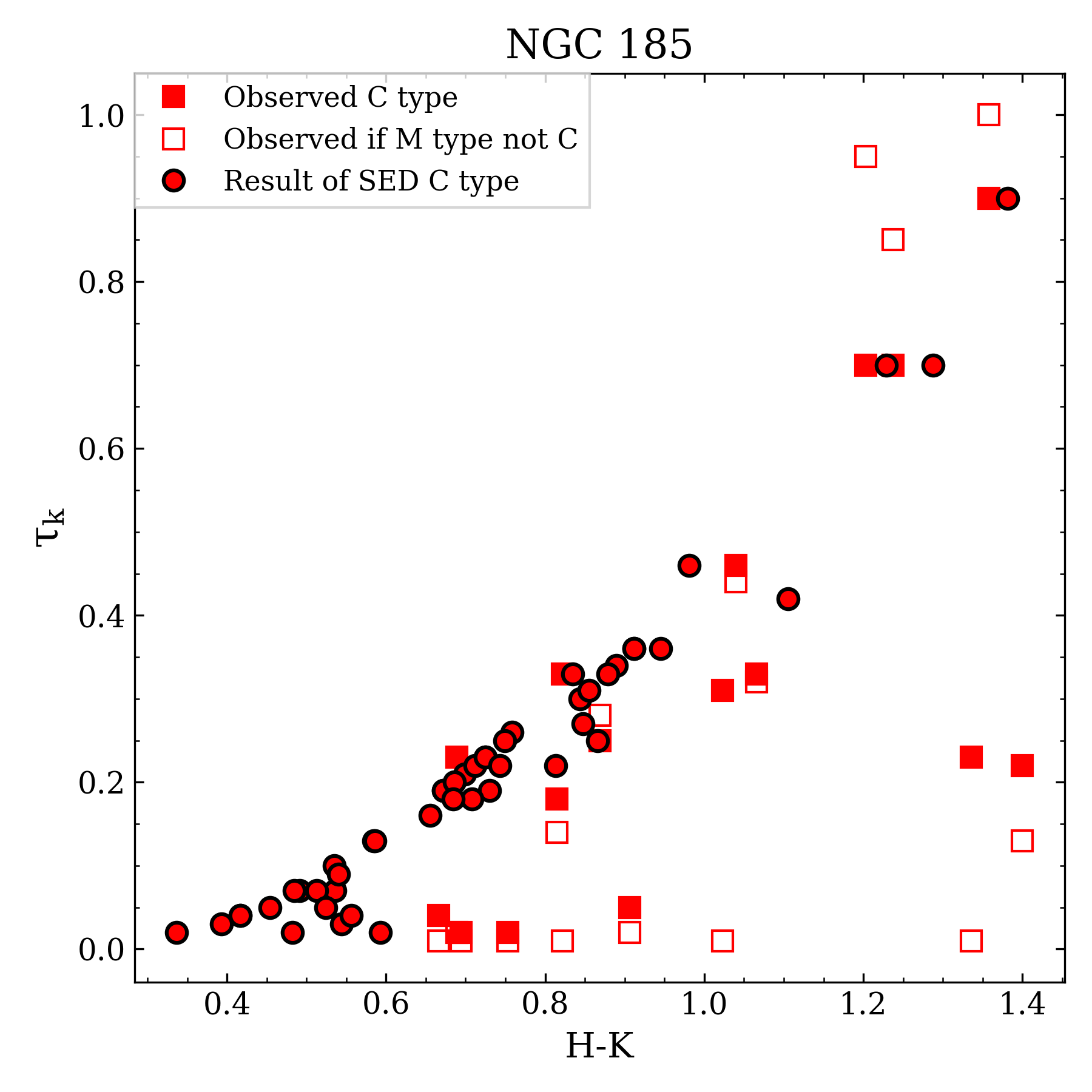}
\caption{ The associations between optical depth in K-band filter and near-infrared colors for carbon stars are represented by red squares, while empty red squares in the plot represent carbon stars that were modeled assuming an oxygen-rich atmosphere. Red circles correspond to carbon stars modeled using synthetic SED calculations. The blue crosses correspond to the results obtained from the \texttt{DESK} code applied to the stars found in the catalogs of \citet{Kang05} and \citet{Sohn06}.}
\label{fig:Tau_K}
\end{center}
\end{figure*}

Figure \ref{fig:Tau_K} plots data from carbon-rich AGB stars in the \citet{Kang05} and \citet{Sohn06} catalogs, modeled using the \texttt{DUSTY} and \texttt{DESK} codes. Filled red squares show \texttt{DUSTY} results for carbon stars, while empty red squares represent carbon star colors with optical depths calculated assuming an oxygen-rich envelope. Blue crosses from \texttt{DESK} show a similar trend for carbon stars but with lower resolution. Circular markers correspond to colors extracted from the best-fitting SEDs for each star.

%In this figure, there are also noticeable deviations from the expected trend. For instance, In the galaxy NGC 147, star ID 3 appears to align more closely with the trend observed for oxygen stars rather than carbon stars. Our calculations, following the method outlined by \citet{Javadi13_III}, further indicate that the initial mass of this particular star was 1.01$\msun$. Consequently, it cannot transition into the regime of carbon stars, providing an explanation for its placement in the oxygen star trend.

Deviations from the expected trends are evident in the Figure \ref{fig:Tau_K}, primarily due to the intrinsic variability of LPVs. Additionally, most of the photometric data were obtained from single-epoch or only a few observations, making it difficult to determine reliable average magnitudes. The absence of time-averaged photometry contributes further to the scatter in the observed colors and optical depths. An example of intrinsic variability can be seen in NGC 147, where star ID 3, a carbon star with a J$−$K color of 3.14 and a K-band optical depth of 0.95, appears closer to the region populated by carbon stars modeled with oxygen-rich dust properties. However, in the H$-$K panel, the same star shows an H$−$K color of 1.85, consistent with other carbon stars.

Similar discrepancies are seen for stars with IDs 47 and 71. Although their best SED is determined based on the chi-square test, the K band in the near-infrared region is not aligned with their SED. Specifically, the reported K band magnitude is higher than expected, resulting in a redder color. However, based on their SED, these stars should be positioned further towards the left of the plot.

Several deviations can be observed in galaxy NGC 185, too. For instance, considering star ID 152 in the \citet{Kang05} catalog, it is reported to have a magnitude of 16.4 mag in the K-band, while in the \citet{Lorenz11} catalog, the same star has a magnitude of 17.1 mag. Upon examining the spectral energy distribution of this star, it becomes apparent that the second magnitude aligns better with the SED. Additionally, for the star with ID 47, the J-band magnitude is lower than expected based on the SED. As a result, this star is shifted towards the right side of the plot.

\subsection{Color}

\begin{figure*}
\begin{center}
\includegraphics[width=0.45\textwidth]{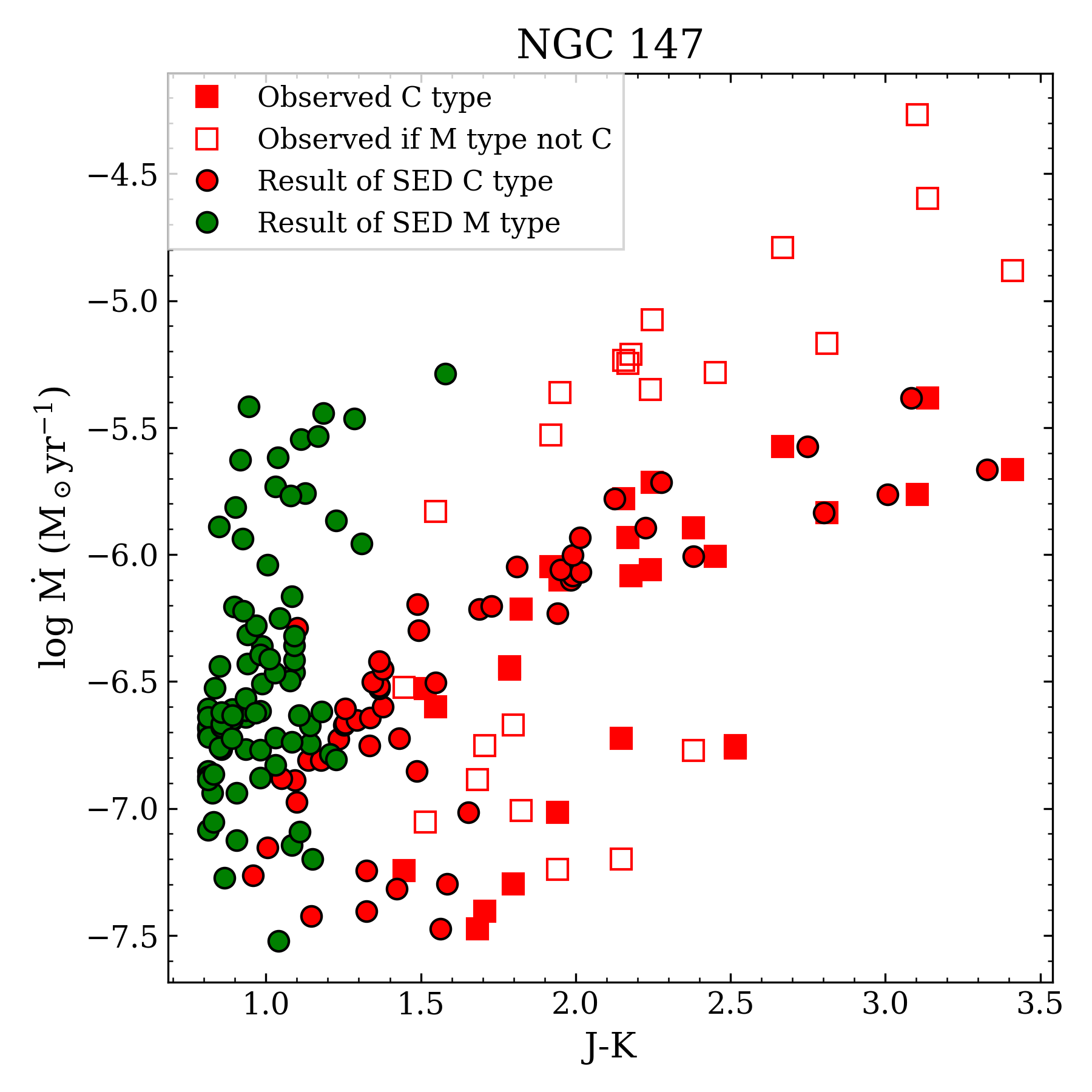}
\includegraphics[width=0.45\textwidth]{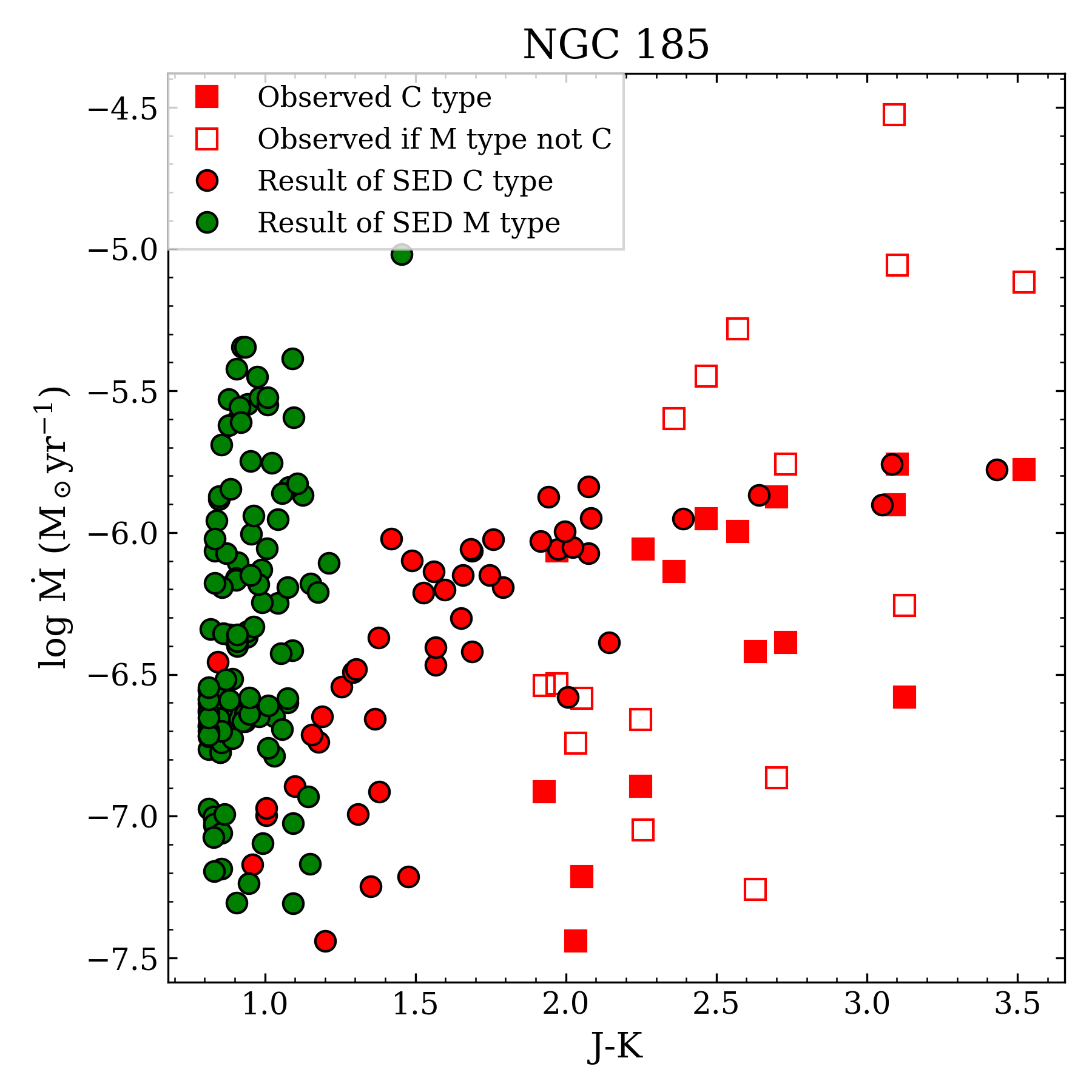}
\caption{ Mass-loss rate as a function of color. It is important to note that the observational data, represented by squares, pertain exclusively to carbon stars. In contrast, the data depicted as circles are derived from the synthetic SED calculations rather than direct observations. }
\label{fig:Mdot_JK}
\end{center}
\end{figure*}

Figure \ref{fig:Mdot_JK} illustrates the correlation between mass-loss rate, log(\textnormal{\.{M}}), and the   J$-$K color index, with observational data available exclusively for carbon stars \citep{Kang05, Sohn06}. As previously noted, the catalogs under consideration contain only carbon stars. To extend the analysis to oxygen-rich LPVs, green circles on the plot represent the derived behavior of oxygen stars, based on SEDs computed using \texttt{DUSTY}.

The J$-$K color serves as a proxy for both the temperature of the stellar atmosphere and the dust content in the circumstellar envelope \citep{Javadi13_III}. Observations indicate that as LPV stars evolve and cool, their J$-$K colors become redder, reflecting the increasing influence of circumstellar dust and molecular absorption on the observed fluxes \citep{Prager22}. This reddening is strongly associated with elevated mass-loss rates, as dust formation is a direct consequence of material being ejected from the stellar atmosphere \citep{Whitelock06, matsuura09}.

Redder J$-$K colors in LPV stars signify cooler, more extended atmospheres conducive to dust formation and accumulation. Once formed, dust grains are propelled outward by radiation pressure, enhancing the mass-loss process. This is further amplified by pulsation-driven shocks, which lift material from the stellar surface to regions where dust can condense \citep{Vassiliadis93}. Consequently, stars with higher mass-loss rates exhibit redder J$-$K colors, driven by the combined effects of lower temperatures and increased optical depth in the circumstellar dust envelope.

As can be seen in the Figure \ref{fig:Mdot_JK}, there is an upward trend in log(\textnormal{\.{M}}) with increasing J$-$K color index. In this figure, only observational data for carbon-rich LPVs were available, and these are shown as red squares. For comparison with the computational results, we extracted the mass-loss rates of all stars in the final catalogs from their best-fitting SEDs. These values are represented by circular markers. Regions associated with C-type LPVs are identified using both the observational data and the SED fits, while regions corresponding to M-type LPVs are determined solely from the SED models. As expected, carbon-rich LPVs occupy regions with more pronounced J$-$K colors than other stars. However, beyond a threshold of J$-$K $>$ 2 mag, this relationship begins to plateau, meaning that further increases in reddening no longer correspond to significant changes in mass-loss rates. This plateau occurs because the enhanced reddening observed in carbon stars at these higher J$-$K values is primarily due to the increasing optical depth of the circumstellar envelope, rather than an actual rise in the amount of material being expelled \citep{Riebel12, Eriksson14}. Additionally, the geometry and distribution of the circumstellar dust can further affect reddening without altering the mass-loss rate.

Interestingly, carbon-rich stars with similar J$-$K colors may not necessarily exhibit higher mass-loss rates than their M-type counterparts. This is because the dust properties in carbon and oxygen-rich stars differ, with carbon stars often having dust compositions that result in stronger reddening effects for a given level of mass-loss. Furthermore, luminosity plays a crucial role in interpreting reddening. Stars with lower luminosities are more prone to reddening due to cooler, denser dust envelopes that scatter and absorb light more effectively. In contrast, more luminous stars, which drive stronger stellar winds, are able to reduce the overall reddening by dispersing their dust more efficiently \citep{vanLoon99}. These complexities underscore the need for caution when interpreting mass-loss rates based on infrared color indices such as J$-$K, particularly at higher values where the relationship becomes nonlinear or inconsistent.

\subsection{Stellar Parameters}

The luminosity (and, as illustrated in Figure \ref{fig:L_MDOT}, mass-loss) in the case of variable stars can be succinctly characterized as the combined influence of two factors: the overall area of the radiating surface and the average surface brightness over that specific area. For spherical stars, the area can be easily computed based on the radius, while the local regulation of surface brightness is determined by the temperature \citep{Freedman10}. Nonetheless, disentangling the impacts of pulsation and temperature on mass-loss is a critical yet challenging endeavor in practice. This is because pulsation intensifies while the temperature diminishes as the star progresses along the RGB, AGB, or red supergiant branch \citep{Whitelock87}. The effects of pulsation amplitude and pulsation period on stellar mass-loss will be discussed in the sections that follow.

\subsubsection{Pulsation period}
\label{sec:period}

Observational studies have been extensively employed over the years to investigate the correlation between mass-loss rates and fundamental pulsation parameters, such as amplitudes and period of fluctuations \citep{Reimers75}. \citet{Hofner18} observes a distinct positive correlation between mass-loss rate and period within the range of 300 to 800 days for semi-regular variables (SRVs). The study reveals that the mass-loss rates of typical Miras and SRVs exhibit an increase with the pulsational period.

%To delve deeper into the mass-loss rate as a function of period, it is crucial to first examine the relationship between period and luminosity. A well-documented positive correlation exists, whereby longer-period variables are typically more luminous. %\citep{Freedman10}. Furthermore, as illustrated in Figure \ref{fig:L_MDOT}, there is a clear positive correlation between mass-loss rate and luminosity. Consequently, the observed positive correlation between mass-loss rate and period is both expected and consistent with these underlying relationships.

This positive correlation is driven by interconnected physical processes involving stellar pulsation, dust formation, and radiation-driven winds. As LPVs evolve along the asymptotic giant branch, their pulsation periods lengthen due to increasing radii and decreasing surface gravities, resulting in stronger radial pulsations \citep{Mowlavi18}. These pulsations generate shock waves that propagate through the extended stellar atmosphere, lifting material to altitudes where temperatures permit dust condensation \citep{Bowen88}. In longer-period LPVs, which are typically more luminous and evolved, enhanced pulsation amplitudes and atmospheric levitation promote dust formation at greater distances from the photosphere. At these distances, radiation pressure efficiently accelerates dust grains and driving higher mass-loss rates \citep{Hofner18}.

\begin{figure*}
\begin{center}
\includegraphics[width=0.45\textwidth]{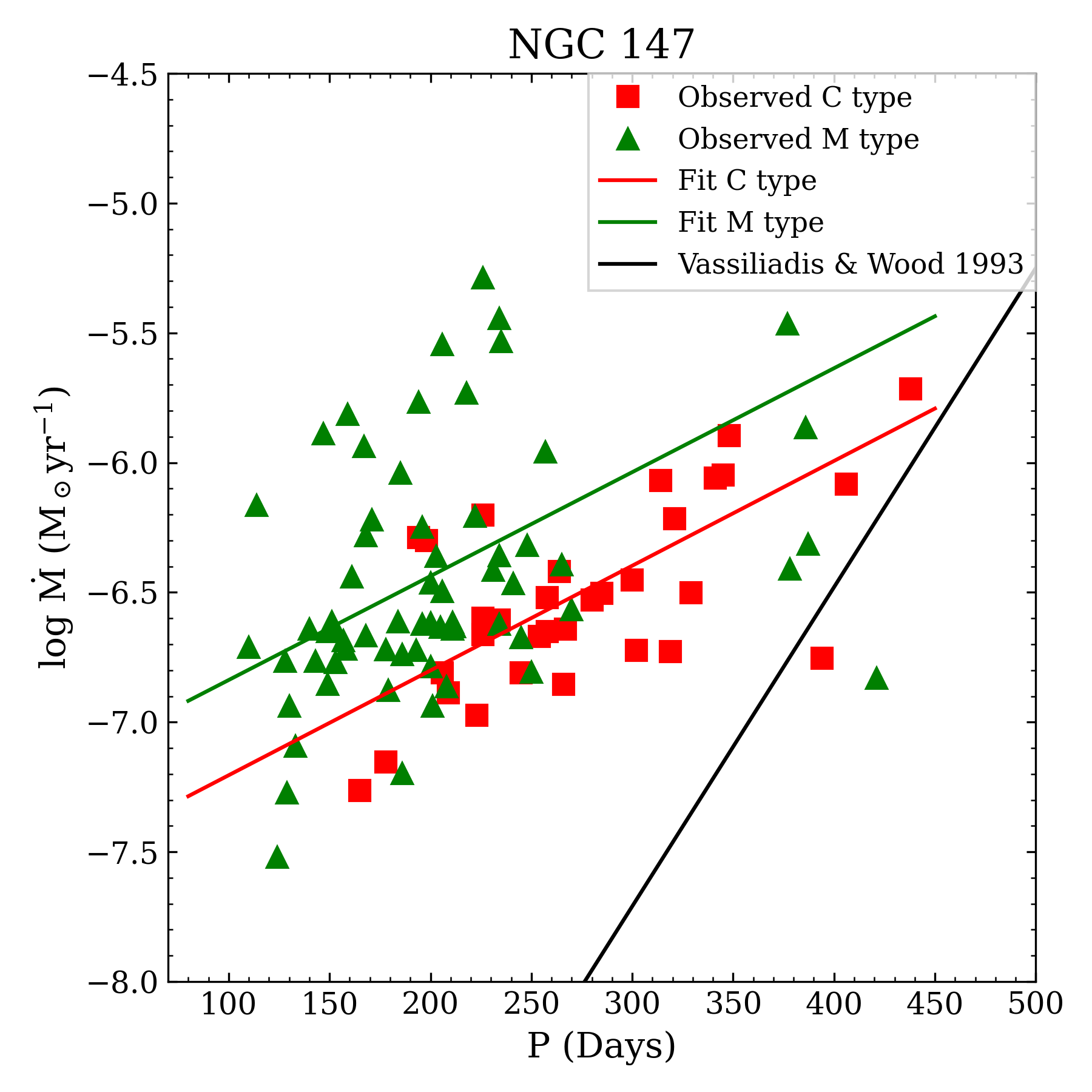}
\includegraphics[width=0.45\textwidth]{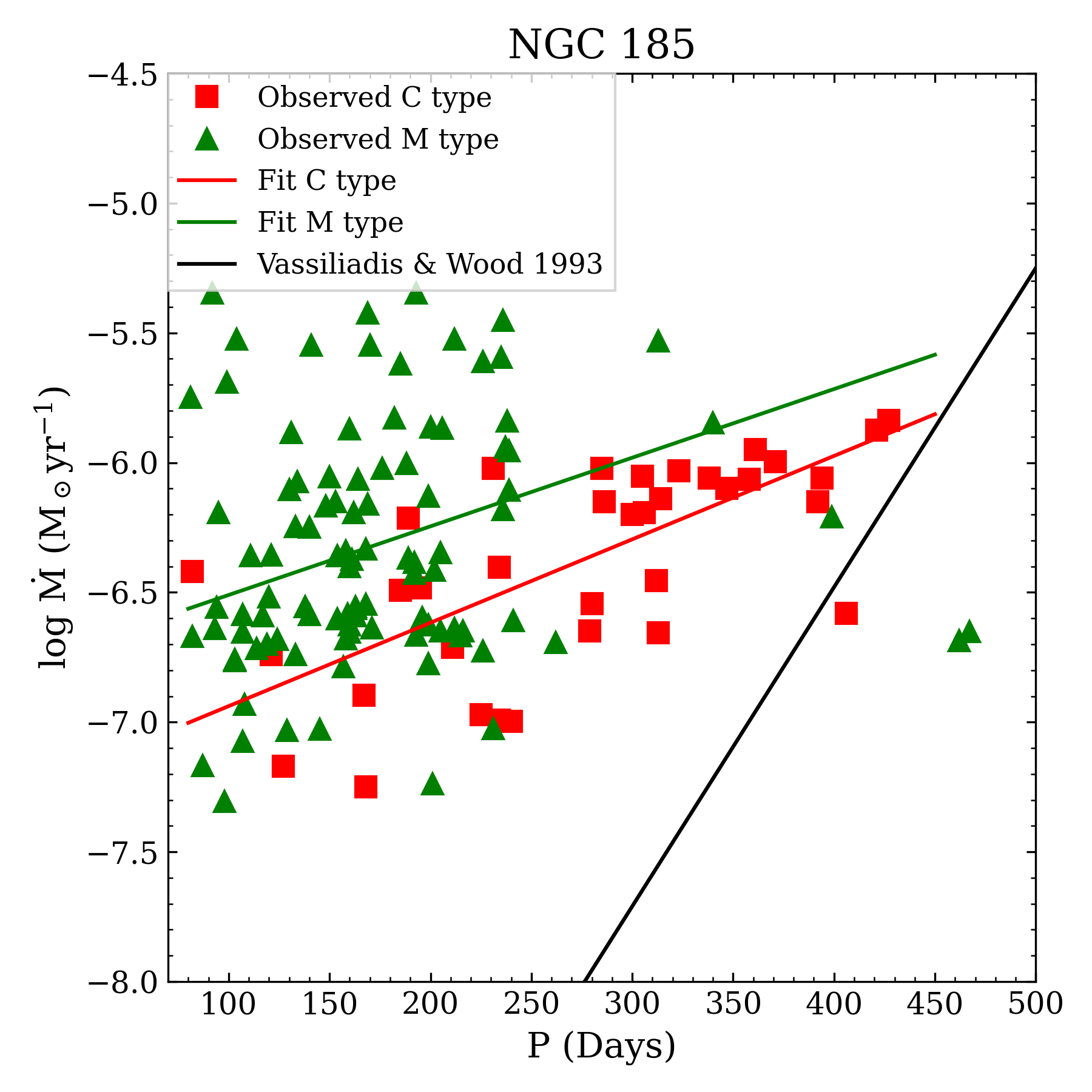}
\caption{ Mass-loss rate as a function of period. Green and red colors represent oxygen- and carbon-rich stars, respectively. The plot showcases the best-fit trends for oxygen LPV stars (green line) and carbon LPVs (red line). For comparison, the black line represents the results reported by \citet{Vassiliadis93}. Refer to Table \ref{tab:amplitude-fit} for the fit parameters. }
\label{fig:Mdot_P}
\end{center}
\end{figure*}

We analyzed the log(\textnormal{\.{M}})-P diagram, which includes all stars for which periods are available \citep{Lorenz11}, as illustrated in Figure \ref{fig:Mdot_P}. Historically, diagrams of this type have been utilized for establishing empirical relationships between period and mass-loss rate, as demonstrated by prior research \citep{Vassiliadis93,Straniero06,Uttenthaler19}.

%\citet{Vassiliadis93}  established a relationship between the periodicity and mass-loss rate of evolved stars. While their study focused on stars with periods exceeding 400 days, they reported a clear positive correlation, expressed as  log\textnormal{\.{M}}(M\textsubscript{\(\odot\)}yr\textsuperscript{-1}) = 0.0123P(days)-11.4. 
%\textbf{Based on our analysis of Figure \ref{fig:Mdot_P}, we derived best-fit relations for NGC 147 and NGC 185, as summarized in Table \ref{tab:amplitude-fit}. For NGC 147, the relations for log\textnormal{\.{M}}(M\textsubscript{\(\odot\)}yr\textsuperscript{-1}) are  0.004(days)-7.238 for M-type stars and  0.004(days)-7.853 for C-type stars. For NGC 185, the relations are 0.003(days)-6.775 for M-type stars and 0.005(days)-7.921 for C-type stars.}

\citet{Vassiliadis93}  established a relationship between the periodicity and mass-loss rate of evolved stars. While their study focused on stars with periods exceeding 400 days, they reported a clear positive correlation, expressed as  log\textnormal{\.{M}}(M\textsubscript{\(\odot\)}yr\textsuperscript{-1}) = 0.0123P(days)-11.4.

% Based on our analysis of Figure \ref{fig:Mdot_P}, we derived best-fit relations for NGC 147 and NGC 185, as summarized in Table \ref{tab:amplitude-fit}. The linear fits for $\log\,\dot{M}$ (in $M_{\odot}\,\text{yr}^{-1}$) as a function of period (in days) are:

% \vspace{1em}
% \textbf{NGC 147:}
% \begin{itemize}
%     \item M-type stars: $\log\,\dot{M} = (0.0040 \pm 0.0014)\,P - (7.2385 \pm 0.0008)$
%     \item C-type stars: $\log\,\dot{M} = (0.0040 \pm 0.0018)\,P - (7.8530 \pm 0.0016)$
% \end{itemize}

% \textbf{NGC 185:}
% \begin{itemize}
%     \item M-type stars: $\log\,\dot{M} = (0.0026 \pm 0.0009)\,P - (6.7747 \pm 0.0003)$
%     \item C-type stars: $\log\,\dot{M} = (0.0048 \pm 0.0013)\,P - (7.9213 \pm 0.0007)$
% \end{itemize}

From our analysis of Figure~\ref{fig:Mdot_P}, we derived similar linear relations for M- and C-type stars in NGC 147 and NGC 185. The best-fit parameters for each population are provided in Table~\ref{tab:amplitude-fit}.

In another study \citet{Hernandez24} provided a detailed analysis comparing Mira and semi-regular variables. They explained the deviations observed between the plot's empirical data and the theoretical trend line. This comparison highlights the differences in mass-loss behaviors among various types of LPV stars, offering insights into the underlying physical mechanisms driving these discrepancies.

Figure \ref{fig:Mdot_P} shows that some stars with periods shorter than 300 days exhibit mass-loss rates exceeding  $10^ {−6} \msun yr^{−1}$ . These stars include IDs 36, 37, 51, 61, 68, 77, 79, and 104 in NGC 147, as well as IDs 9, 14, 64, 68, 87, 89, 90, 92, 93, 96, 119, 120, 127, 128, 132, 140, 146, and 158 in NGC 185. Examining their SEDs in Figures \ref{fig:SED_fits_147} and \ref{fig:SED_fits_185} reveals that all of these stars exhibit prominent emission at wavelengths beyond the mid-infrared. This suggests that they contain a higher dust content compared to other LPVs.

%In Figure \ref{fig:Mdot_P}, stars with periods exceeding 200 days and mass-loss rates below $10^{-7} \msun yr^{-1}$ exhibit observed magnitudes in the K-band that deviate from the best-fit SED derived from their observational data. In NGC 147, these stars are identified as IDs 41, 45, 48, 50, 53, 66, 113, and 116, while in NGC 185, they correspond to IDs 42, 75, 91, 106, 109, 111, and 115.

%\clearpage

{\small % Adjust the font size to make the table smaller
\setlength{\tabcolsep}{2.2pt} % Adjust column spacing
\begin{table}
\begin{center}
 \caption{  Best-fit parameters derived from Figure \ref{fig:Mdot_P} for the galaxies NGC 147 and NGC 185. The parameters a and b correspond to the slope and intercept, respectively, of the relation log\textnormal{\.{M}}(M\textsubscript{\(\odot\)}yr\textsuperscript{-1}) = a$\times$P(days)+b. }
% \resizebox{\columnwidth}{!}{
\begin{tabular}{cccc}
\hline
\hline
Galaxy                   & Chemical type & a & b \\ \hline
\multirow{2}{*}{NGC 147} & M-type        &  0.0040 $\pm$ 0.0014 &  $-$7.2385 $\pm$ 0.0008  \\ \cline{2-4} 
                         & C-type        &  0.0040 $\pm$ 0.0018 &  $-$7.8530 $\pm$ 0.0016 \\ \hline
\multirow{2}{*}{NGC 185} & M-type        &  0.0026 $\pm$ 0.0009 & $-$6.7747 $\pm$ 0.0003  \\ \cline{2-4} 
                         & C-type        &  0.0048 $\pm$ 0.0013  & $-$7.9213 $\pm$ 0.0007  \\ \hline
\hline
\end{tabular}
%}
\label{tab:amplitude-fit}
\end{center}
\end{table}
}

\subsubsection{Pulsation amplitude}

%\textbf{Stellar pulsations are a primary driver of mass-loss in stars \citep{McDonald16}. As the amplitude of these pulsations increases, the rate of mass-loss is expected to rise correspondingly. Figure~\ref{fig:Mdot_Amp} shows the relationship between mass-loss and pulsation amplitude in the i-band. While the data suggest a weak positive trend, with slopes of $0.14 \pm 0.12$ for NGC 147 and $0.10 \pm 0.07$ for NGC 185, the scatter is considerable, making the correlation difficult to quantify precisely.}

Stellar pulsations drive mass-loss in stars \citep{McDonald16}. Higher pulsation amplitudes generally lead to increased mass-loss rates. For LPV stars, the relationship between pulsation amplitude and mass-loss rate is positive but more complex than the period-mass-loss correlation discussed in Section \ref{sec:period}. In Mira-type variables, large pulsation amplitudes create strong atmospheric shock waves that lift gas to higher altitudes, where cooler temperatures allow dust to form \citep{Bowen88,Hofner18}. This process enhances dust grain formation, enabling radiation pressure to drive stellar winds more effectively. As a result, stars with larger pulsation amplitudes experience greater dust-driven mass-loss due to the extended atmospheres, which facilitate efficient acceleration of dust and gas.

Observations support this mechanism. Mira variables, with visual amplitudes typically exceeding 2.5 magnitudes in V-band, show strong infrared excesses and high mass-loss rates compared to semi-regular variables with lower amplitudes \citep{Whitelock08}. Additionally, dynamic atmosphere simulations confirm that higher pulsation amplitudes enhance dust formation and wind acceleration, supporting observations of increased mass-loss \citep{Winters2000}.

\begin{figure*}
\begin{center}
\includegraphics[width=0.45\textwidth]{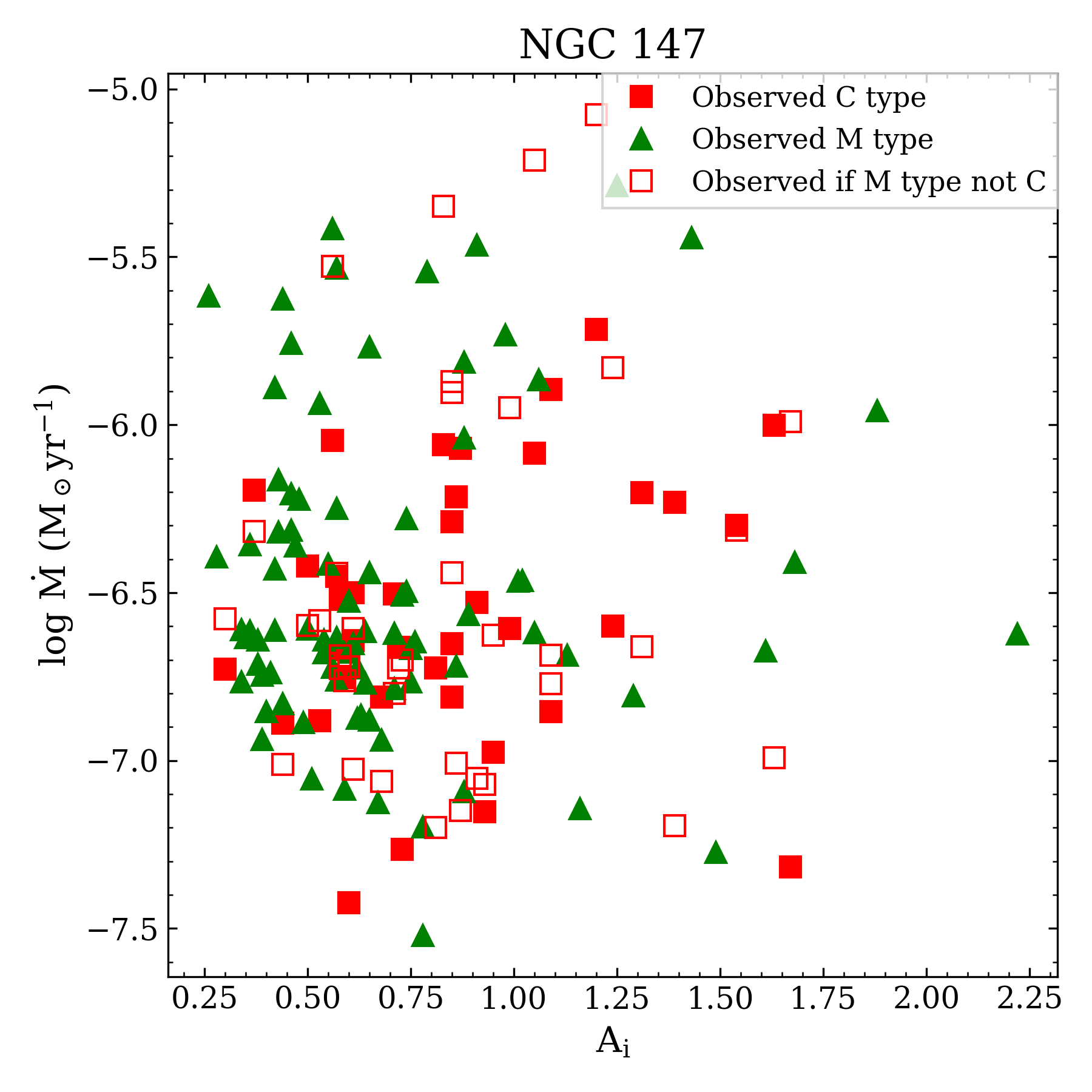}
\includegraphics[width=0.45\textwidth]{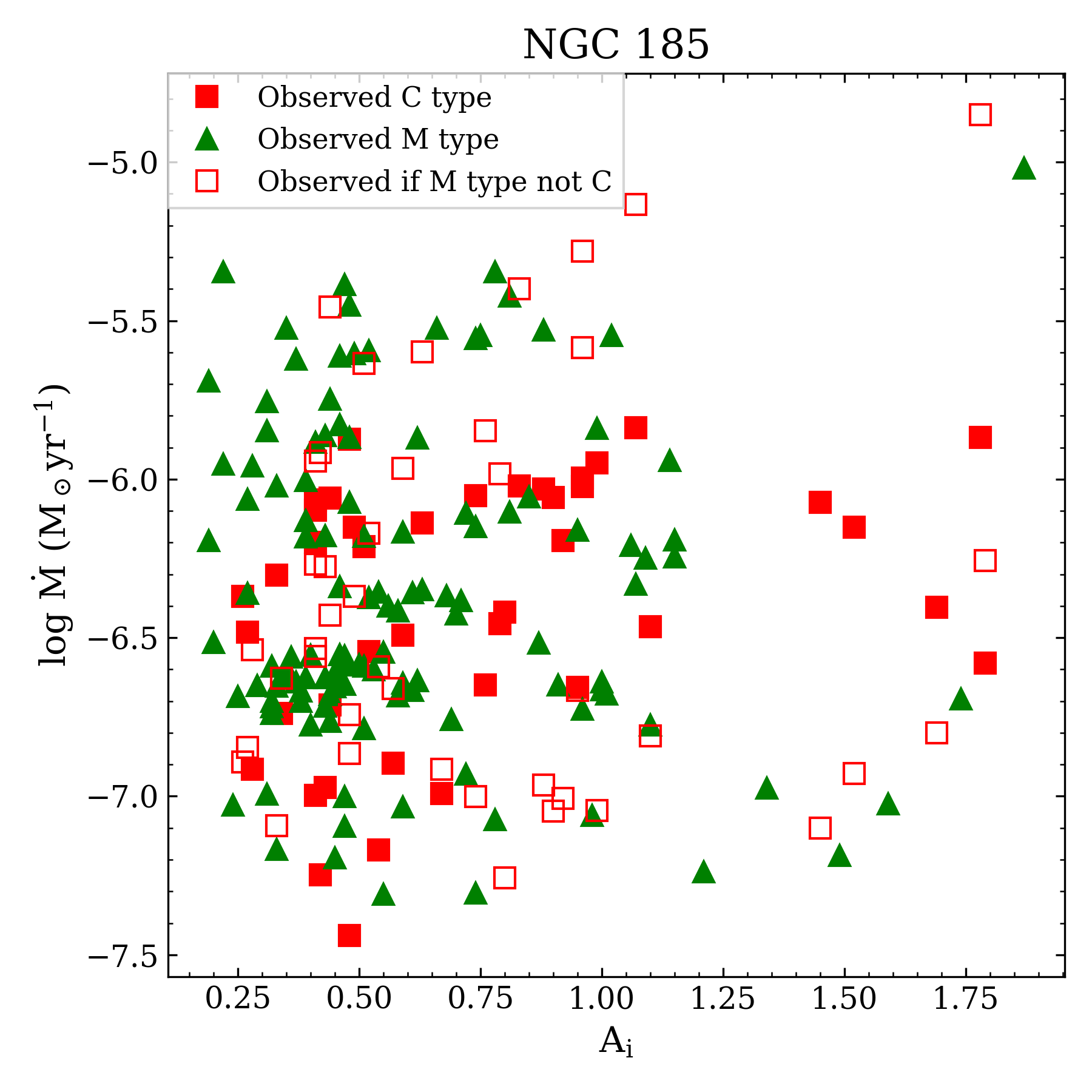}
\caption{ Mass-loss rate as a function of amplitude in the i-filter band. Oxygen-rich stars are represented by the color green, while carbon-rich stars are depicted in red. Amplitudes are adopted from \citet{Lorenz11}.}
\label{fig:Mdot_Amp}
\end{center}
\end{figure*}

Figure~\ref{fig:Mdot_Amp} shows the relationship between mass-loss and pulsation amplitude in the i-band. While the data suggest a weak positive trend, with slopes of $0.14 \pm 0.12$ for NGC 147 and $0.10 \pm 0.07$ for NGC 185, the scatter is considerable, making the correlation difficult to quantify precisely.

The observed amplitude variations are influenced by the logarithmic nature of the magnitude scale. As a result, a star may be intrinsically bright (and thus exhibit a high mass-loss rate, as shown in Figure~\ref{fig:L_MDOT}) and still undergo substantial variability, yet these changes may appear modest when expressed in magnitudes. This helps explain the presence of stars with low apparent amplitudes but high mass-loss rates in Figure~\ref{fig:Mdot_Amp}. In contrast, luminosity provides a more direct and linear representation of stellar variability, revealing a clearer relationship between pulsation amplitude and mass-loss rate.

Furthermore, measuring pulsation amplitudes in the K-band is recommended due to its reduced sensitivity to stellar temperature variations, circumstellar extinction, and dust emission. The K-band lies near the peak of the SED of LPV stars and offers an optimal balance between attenuation at shorter wavelengths and emission at longer wavelengths. This makes it especially suitable for characterizing variability and deriving accurate bolometric corrections.

%@@@@@@@@@@@@@@@@@@@@@@@@@@@@@@@@@@@@@@@@@@@@@@@@@@@@@@@@@@@@@@@@@@@@@@@@@@@@@@@@@@@@@@@@@@

\section{Discussion}
\label{sec:discussion}

%In this study, we calculated the mass-loss rates of long-period variable stars in the dwarf galaxies NGC 147 and NGC 185. By cross-matching these stars with multiple astronomical catalogs, we obtained magnitudes across several photometric bands. Through spectral energy distribution fitting, we derived critical physical parameters, including luminosity, effective temperature, and mass-loss rates. Our analysis allowed us to distinguish between carbon-rich and oxygen-rich atmospheres, which significantly influence the mass-loss processes of these stars.

In this section, we assess the implications of the derived mass-loss rates for LPV and AGB stars in the dwarf galaxies NGC 147 and NGC 185, with particular emphasis on their role in the galactic dust budget. Using multi-wavelength photometric data combined with spectral energy distribution fitting, we quantify the contributions of carbon-rich and oxygen-rich stars to the ISM and compare these results with those reported in previous studies. We further examine the spatial distribution of mass-losing stars within each galaxy to identify structural differences that may reflect variations in their stellar populations or evolutionary histories.

\subsection{Mass-Loss Rates of AGBs}
\label{sec:ML_AGBs}

In the previous sections, we estimated the mass-loss rates for LPV stars with available data in at least two distinct spectral regions, such as the visible, near-infrared, and mid-infrared. However, this subset represents only a fraction of the total AGB population. Since a significant portion of the dust and gas expelled by AGB stars originates from sources not included in our final LPV catalogs (primarily due to the lack of multi-wavelength photometry) it is essential to account for the full AGB population when estimating the total dust input into the interstellar medium. 

% To estimate the total mass-loss rate of AGB stars in NGC 147 and NGC 185, we consider the individual contributions of AGB stars. As described in Section~\ref{sec:catalogs}, the catalog compiled by \citet{Nowotny03}, encompassing stars observed in the V and i magnitude filters, provides the most comprehensive dataset for AGB stars in these galaxies. It includes 146 C-type and 950 M-type stars in NGC 147, and 154 C-type and 1732 M-type stars in NGC 185. The estimation process involves three key steps:

\begin{figure*}
\begin{center}
\includegraphics[width=0.45\textwidth]{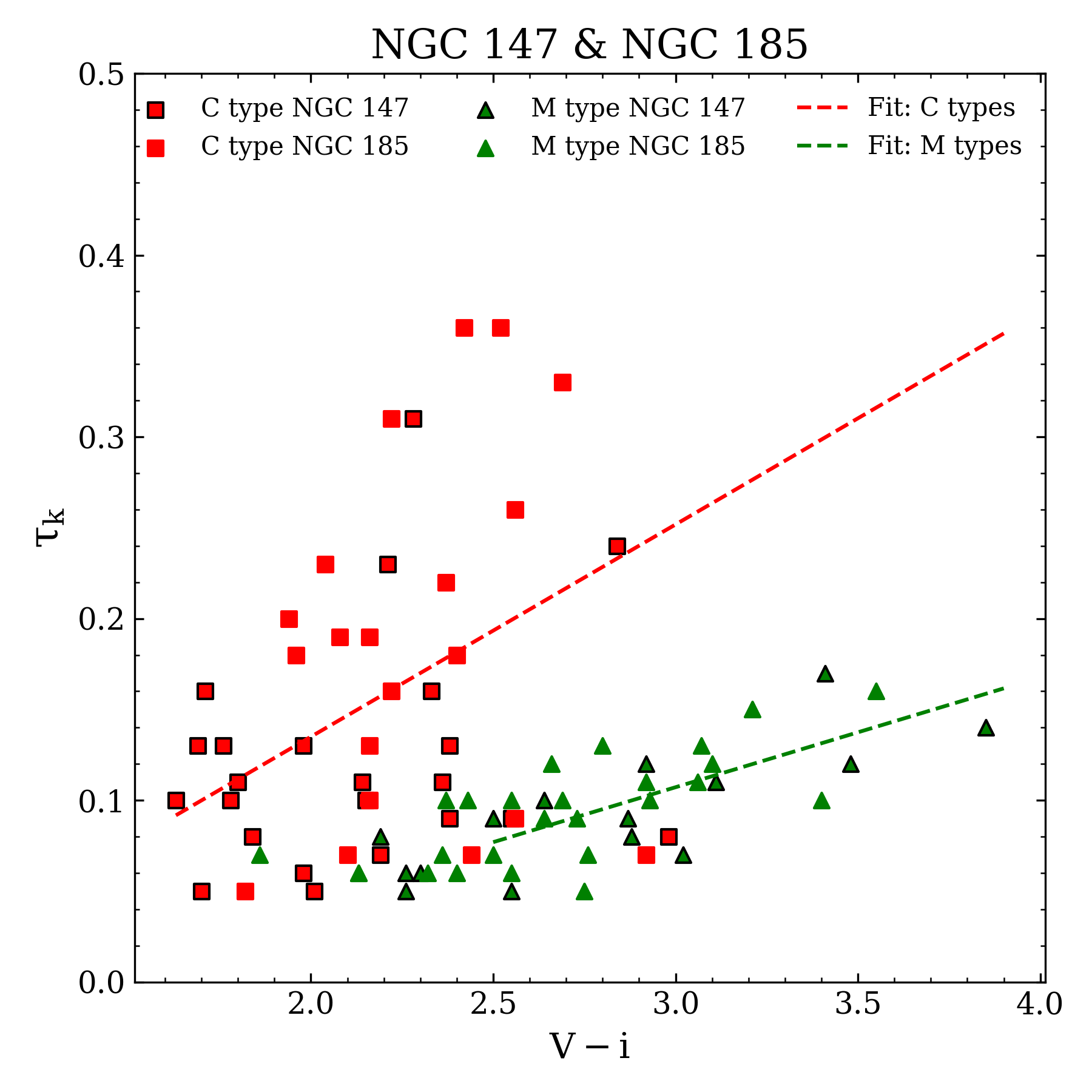}
\includegraphics[width=0.45\textwidth]{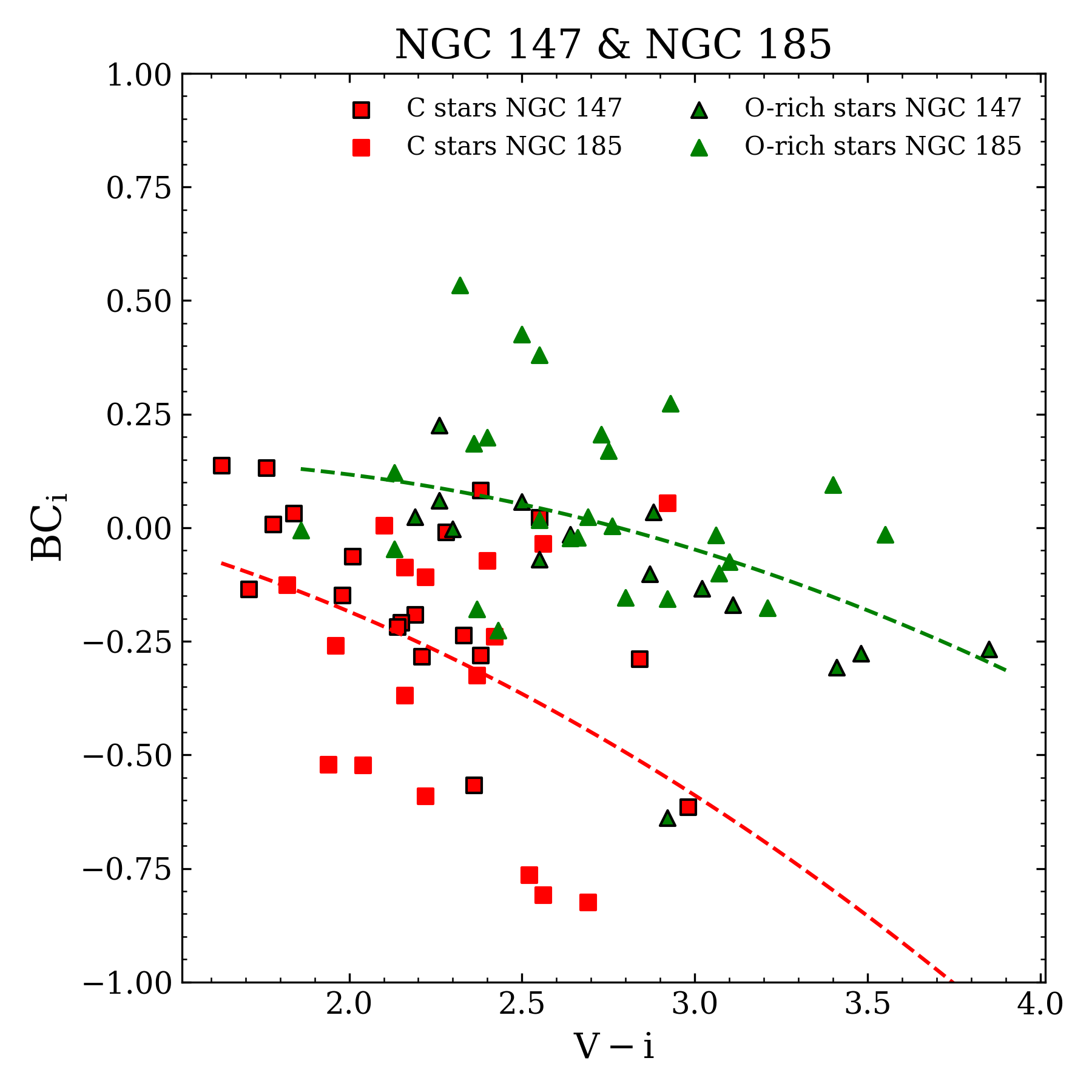}
\caption{Bolometric correction in the i-band vs. color, and optical depth in the K-band vs. color, for stars in NGC 147 and NGC 185. In this illustration, carbon stars are represented by the red squares, while M-type stars are denoted by the green triangles. The plot also illustrates the best-fit trends for oxygen LPV stars, represented by the green dashed-line, and carbon LPVs, shown by the red dashed-line. The parameters of the fitted curves are listed in Table \ref{tab:Fits_Vi}.}
\label{fig:BC_vi}
\end{center}
\end{figure*}

To estimate the total mass-loss rate of AGB stars in NGC 147 and NGC 185, we consider the individual contributions of confirmed AGB stars. As described in Section~\ref{sec:catalogs}, we use the catalog compiled by \citet{Nowotny03}, which is based on photometry in the V and i filters, along with additional narrow-band TiO and CN observations to identify AGB stars. While the full catalog includes 18300 stars in NGC 147 and 26496 stars in NGC 185, only a subset of these are AGB stars. Specifically, \citet{Nowotny03} identified 146 carbon-rich (C-type) and 950 oxygen-rich (M-type) AGB stars in NGC 147, and 154 C-type and 1732 M-type AGB stars in NGC 185. The estimation of the total mass-loss rate involves three key steps: (1) calculating the optical depth of the AGB stars based on established relationships between optical depth and color (which are described in detail in the subsequent text and illustrated in Figure~\ref{fig:BC_vi}, with the corresponding fitting parameters listed in Table~\ref{tab:Fits_Vi}),  (2) deriving the luminosity of each AGB star using bolometric corrections, and (3) computing the mass-loss rates using Equation~\ref{eq:mdot_const}.

\begin{table}
\begin{center}
\caption{A list of the parameters of the fitted curves in the Figure\ref{fig:BC_vi}. }
\begin{tabular}{cccc}
\hline
\hline
\multicolumn{4}{c}{$\tau_{k} = a \times (V-i) + b$}     \\ \hline
            &    & a          & b                      \\ \hline
M-type      &    &   0.060 $\pm$ 0.014         &   $-$0.074 $\pm$ 0.041                    \\ \hline
C-type      &    &   0.117 $\pm$ 0.052         &   $-$0.099 $\pm$ 0.108                     \\ \hline
\hline
\multicolumn{4}{c}{{$BC_{i} = A \times (V-i)^{2} + B \times (V-i)+ C$}} \\ \hline
                 & A          & B          & C          \\ \hline
M-type           &     $-$0.068 $\pm$ 0.129       &   0.177 $\pm$ 0.732         &    0.037 $\pm$ 1.025        \\ \hline
C-type           &    $-$0.085 $\pm$ 0.358        &   0.019 $\pm$ 1.653         &    0.116 $\pm$ 1.890        \\ \hline
\hline
\end{tabular}
\label{tab:Fits_Vi}
\end{center}
\end{table}

To derive the relationship between optical depth and the V$-$i color of AGB stars, and bolometric corrections, stars classified as carbon-rich or oxygen-rich in the \citet{Nowotny03} catalog were selected from  our final dataset. A key consideration in using this catalog to estimate mass-loss rates is that the observations were conducted at a single epoch. Although single-epoch data can be used to establish correlations between bolometric correction and color, such measurements introduce significant scatter, potentially leading to inaccuracies. As demonstrated by \citet{Whitelock08}, variability in the magnitudes of AGB stars (particularly in the V band) can exceed 2.5 magnitudes. 

It’s worth mentioning that we  excluded stars with optical depths below 0.01, since this is the lower limit of our model grid and values below it are within the measurement uncertainties. In other words, these low values cannot be reliably distinguished from zero. Furthermore, given that the distances to NGC 147 and NGC 185 are approximately equal within the uncertainties (see Table~\ref{tab:quantities}), data from both galaxy catalogs were combined to establish the relationship used for estimating the mass-loss rates of AGB stars. The resulting fits are presented in Figure~\ref{fig:BC_vi} and Table~\ref{tab:Fits_Vi}.

Our calculations indicate that the total mass-loss rate from AGB stars is \(9.44 \times 10^{-4}\ \msun\,\mathrm{yr}^{-1}\) for NGC~147 and \((1.58  \times 10^{-3}\ \msun\,\mathrm{yr}^{-1}\) for NGC~185. The contributions from carbon-rich and oxygen-rich stars, along with their associated uncertainties, are provided in Table~\ref{tab:CMmassloss}. 

\begin{table}[h!]
\centering
\caption{Mass-loss rates from C-type and M-type stars in NGC~147 and NGC~185.}
\label{tab:CMmassloss}
\begin{tabular}{@{}lccc@{}}
\toprule
Galaxy & Star Type & Mass-loss Rate \([\msun\,\mathrm{yr}^{-1}]\) &  [\%] \\ 
\hline
\multirow{2}{*}{NGC~147} 
 & C-type & \((3.86 \pm 1.07) \times 10^{-5}\) & 4.1 \\
 & M-type & \((9.06 \pm 3.78) \times 10^{-4}\) & 95.9 \\
 & Total & \((9.44 \pm 3.78) \times 10^{-4}\) & 100.0 \\
\hline
\multirow{2}{*}{NGC~185} 
 & C-type & \((5.76 \pm 1.60) \times 10^{-5}\) & 3.6 \\
 & M-type & \((1.52 \pm 0.63) \times 10^{-3}\) & 96.4 \\
 & Total & \((1.58 \pm 0.63) \times 10^{-3}\) & 100.0 \\
\hline
\hline
\end{tabular}
\end{table}

\subsection{Dust}
\label{sec:dust}

To estimate the contribution of AGB stars to the galactic dust budget, we apply a gas-to-dust mass ratio of 160 \citep{Marleau10} to the total AGB mass-loss rates derived in Section~\ref{sec:ML_AGBs}. This yields dust injection rates of \(5.90  \times 10^{-6}\ \msun\ \mathrm{yr}^{-1}\) for NGC 147 and \(9.87 \times 10^{-6}\ \msun\ \mathrm{yr}^{-1}\) for NGC 185 (see Table~\ref{tab:dust_summary} for details and uncertainties).

For comparison, \citet{DeLooze16} reported lower dust injection rates based on 65 and 103 carbon-rich AGB stars in NGC 147 and NGC 185, respectively. Their oxygen-rich AGB estimates were scaled from LMC ratios, yielding even smaller values. In contrast, our study directly accounts for both C-type and M-type stars based on the \citet{Nowotny03} catalog, without relying on external scaling relations.

Assuming a dust life time of $\sim$1 Gyr \citep{DeLooze16}, the accumulated dust mass from AGB stars reaches $5.90  \times 10^{3}$ M${\odot}$ in NGC 147 and $9.87 \times 10^{3}$ M${\odot}$ in NGC 185. The value for NGC 185 agrees with the observed dust mass from \textit{Herschel} ($5.1 \times 10^3$ M$\odot$), while for NGC 147, the predicted dust mass far exceeds the observed 128 M$\odot$, suggesting efficient dust removal or destruction processes.

Additional measurements by \citet{Marleau10} and \citet{Temi04} support these conclusions (Table~\ref{tab:dust_summary}). In particular, NGC 185 consistently shows a dust content compatible with AGB injection rates, while NGC 147 remains dust-poor despite a comparable mass-loss rate, potentially due to environmental effects such as tidal stripping or outflows.

\begin{deluxetable*}{lccl}

\tablecaption{Dust injection and dust mass estimates for NGC~147 and NGC~185.\label{tab:dust_summary}}
\tablehead{
\colhead{Quantity} & \colhead{NGC 147} & \colhead{NGC 185} & \colhead{Notes}
}
\startdata
Total dust injection rate [$\msun yr^{-1}$]  & $(5.90 \pm 2.36) \times 10^{-6}$ & $(9.87 \pm 3.97) \times 10^{-6}$ & This study,  \\
& & & gas-to-dust ratio = 160 \\
Dust injection rate (C-type only) [$\msun yr^{-1}$]  & $1.22 \times 10^{-7}$ & $1.93 \times 10^{-7}$ & \citep{DeLooze16} \\
Dust injection rate (M-type, scaled) [$\msun yr^{-1}$] & $4.9 \times 10^{-8}$ & $7.8 \times 10^{-8}$ & Estimated via LMC scaling \\
 & & & \citep{DeLooze16} \\
Accumulated dust over 1 Gyr [$\msun$] & $(5.90 \pm 2.36) \times 10^{3}$ & $(9.87 \pm 3.97) \times 10^{3}$ & This study \\
Observed dust mass [$\msun$] & 128 & $5.1 \times 10^{3}$ &  Herschel \citep{DeLooze16} \\
Observed dust mass (cold dust) [$\msun$]& $\leq 4.0 \times 10^2$ & $1.6 \times 10^{3}$ &  $T_{\text{dust}} \sim 20$ K \citep{Temi04} \\
Total injected (upper limit) [$\msun$] & $4.5 \times 10^2$ & $1.9 \times 10^{3}$ & \citep{Marleau10} \\
\enddata
%\tablecomments{Table formatted to fit within the ApJ two-column layout.}
\end{deluxetable*}

\subsection{Mass-Loss Rates of LPVs}

\label{sec:ML_LPVs}

We estimated mass-loss rates for LPV stars in NGC 147 and NGC 185 based on cross-matched data from the \citet{Lorenz11} and \citet{Nowotny03} catalogs, using the \texttt{DUSTY} radiative transfer code where applicable. In NGC 147, rates were derived for 163 out of 213 LPVs, and in NGC 185 for 187 out of 513. For stars without direct measurements, we applied the method described in Section \ref{sec:ML_AGBs}.

Table~\ref{tab:ml_summary} summarizes the total and individual mass-loss rates. The total LPV mass-loss is $7.8 \times 10^{-5}\msun\mathrm{yr}^{-1}$ for NGC 147 and $3.0 \times 10^{-4}\msun\mathrm{yr}^{-1}$ for NGC 185. These values are expressed as a fraction of the total mass of all AGB stars (with LPVs included). Scaling these values to the broader AGB population, based on the relative catalog sizes (see Table~\ref{tab:ml_summary}), suggests proportional consistency.

Mass-loss rates span roughly $10^{-8}$ to $10^{-6}\msun\mathrm{yr}^{-1}$ in both galaxies, consistent with values reported for LPVs in other M31 satellites. For example, \citet{Abdollahi23} found rates ranging from $1.7 \times 10^{-7}$ to $1.9 \times 10^{-5}\msun\mathrm{yr}^{-1}$ in And-IX, while \citet{Saremi20} reported an average of $2.8 \times 10^{-6}\msun\mathrm{yr}^{-1}$ in And-I. This suggests similar mass-loss behavior among AGB stars across the Local Group satellites.

\begin{table*}[ht]
\centering
\caption{LPV mass-loss summary}
\label{tab:ml_summary}
\begin{tabular}{lcc}
\hline
\hline
 & \textbf{NGC 147} & \textbf{NGC 185} \\
\hline
$N_\mathrm{LPV}$ (used / total) & 163 / 213 & 187 / 513 \\
$N_\mathrm{AGB}$ (total) & 1096 & 1886 \\
$\dot{M}_\mathrm{tot}$ ($\msun\,\mathrm{yr}^{-1}$) & $(7.8 \pm 3.1)\times10^{-5}$ & $(3.0 \pm 1.2)\times10^{-4}$ \\
$\dot{M}_\mathrm{range}$ ($\msun\,\mathrm{yr}^{-1}$) & [$(1.0 \pm0.2) \times 10^{-8}$ , $(5.1 \pm 2.1) \times 10^{-6} $] & [$(1.0 \pm0.2) \times 10^{-8} $ , $(9.5 \pm 3.8) \times 10^{-6} $] \\
\hline
\hline
\end{tabular}
\end{table*}

 \subsection{Spatial Distribution}

\begin{figure*}
\begin{center}
\includegraphics[width=0.45\textwidth]{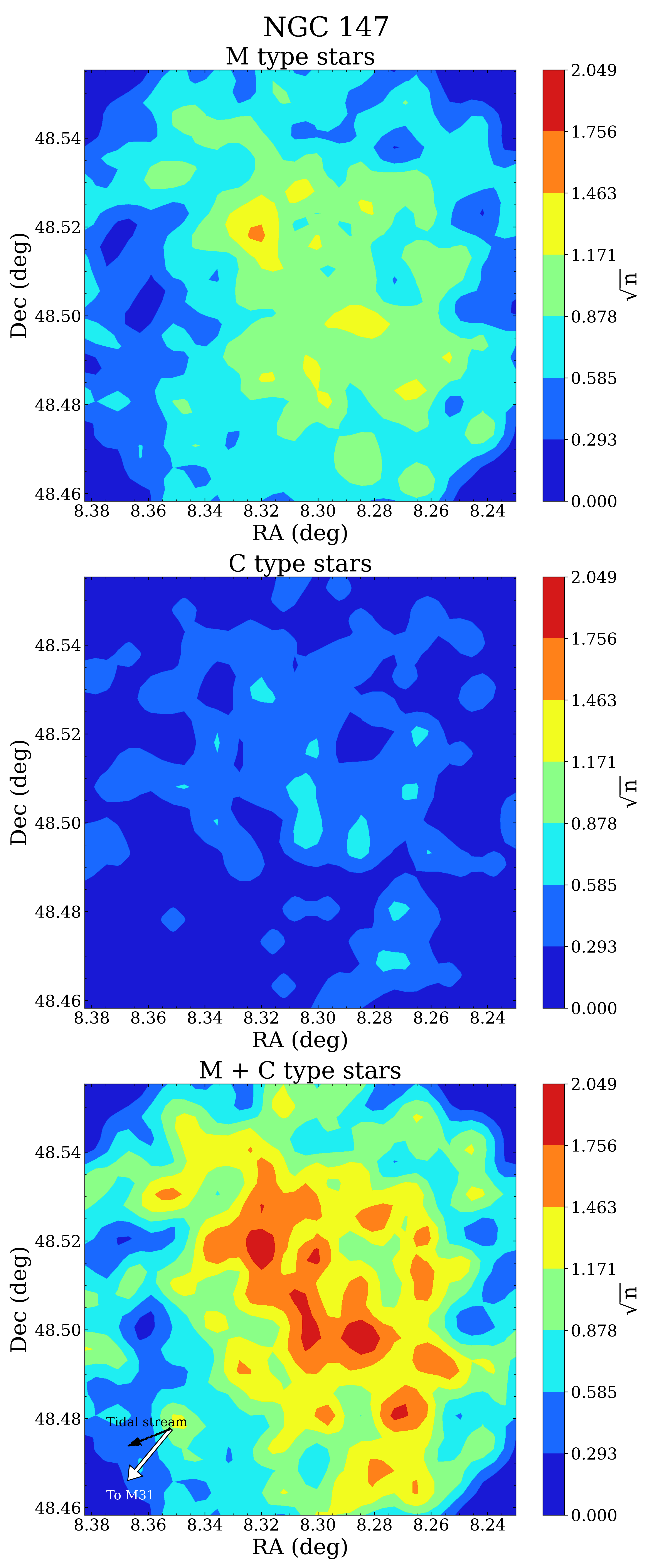}
\includegraphics[width=0.45\textwidth]{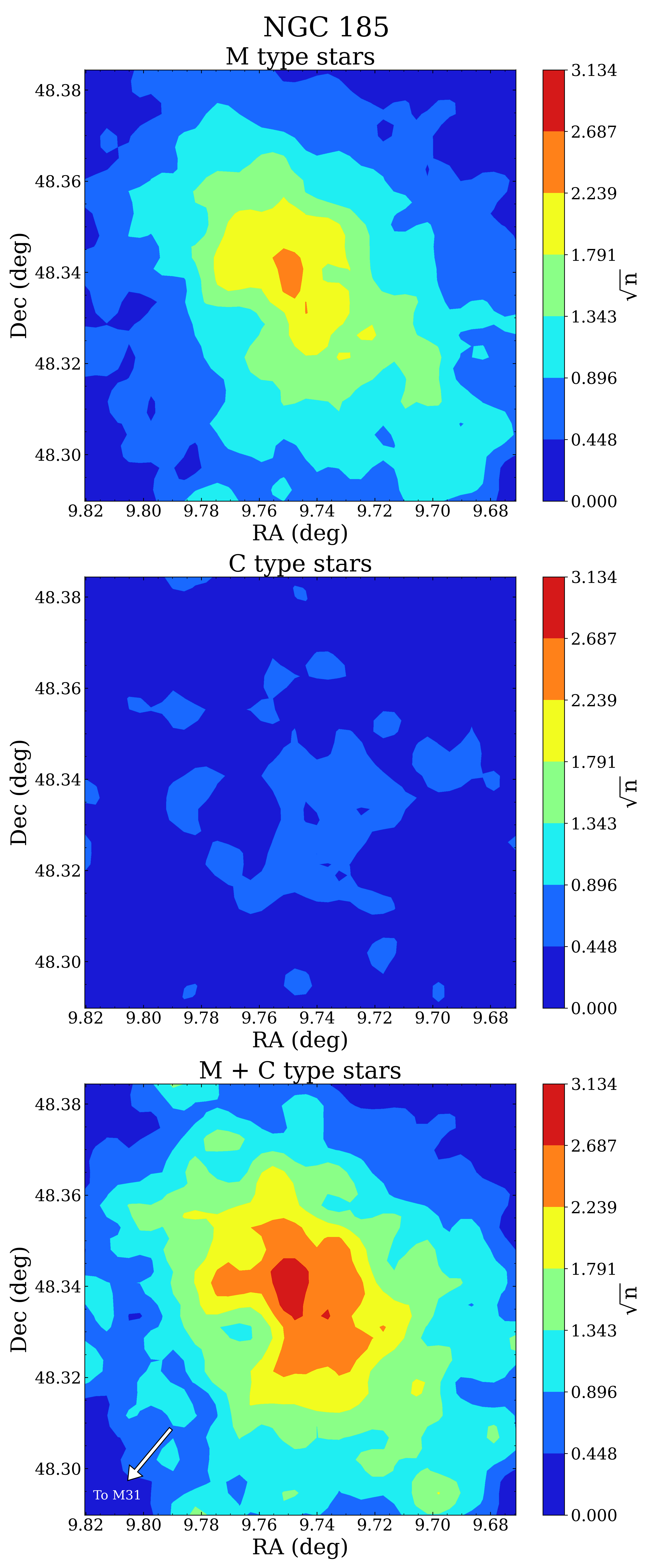}
\caption{ Spatial distribution maps of number surface density of carbon-rich (C-type), oxygen-rich (M-type), and combined (M+C-type) stars within the galaxies NGC 147 and NGC 185, displayed individually. White arrows indicate the approximate direction toward the center of M31, while the black arrow (shown only for NGC 147) indicates the approximate direction of the tidal stream, as adapted from \citet{Preston25}.}
\label{fig:C_M_Contour}
\end{center}
\end{figure*}

The spatial distribution of AGB stars reveals marked differences between NGC 185 and NGC 147 (Figure~\ref{fig:C_M_Contour}). In NGC 185, both M-type and C-type stars show strong central clustering, forming a compact core. In contrast, NGC 147 exhibits a fragmented structure with multiple density peaks and a more dispersed stellar distribution.

Quantitative measures of the distributions are summarized in Table~\ref{tab:spatial_stats}, including star counts, central coordinates, radial dispersions, and Kolmogorov–Smirnov (K-S) test results. For M-type stars, the K-S test indicates a statistically significant difference in radial distributions between NGC 147 and NGC 185 ($P = 0.00$), reflecting the more fragmented, multi-peaked structure in NGC 147 compared to the centralized clustering in NGC 185. In contrast, the radial distributions of C-type stars are not significantly different ($P = 0.40$). The lack of significance for C-type stars may be attributed to the smaller sample sizes, which likely reduce the statistical power of the test, as well as the comparable radial dispersions between the two galaxies.

\begin{figure*}
\begin{center}
\includegraphics[width=0.45\textwidth]{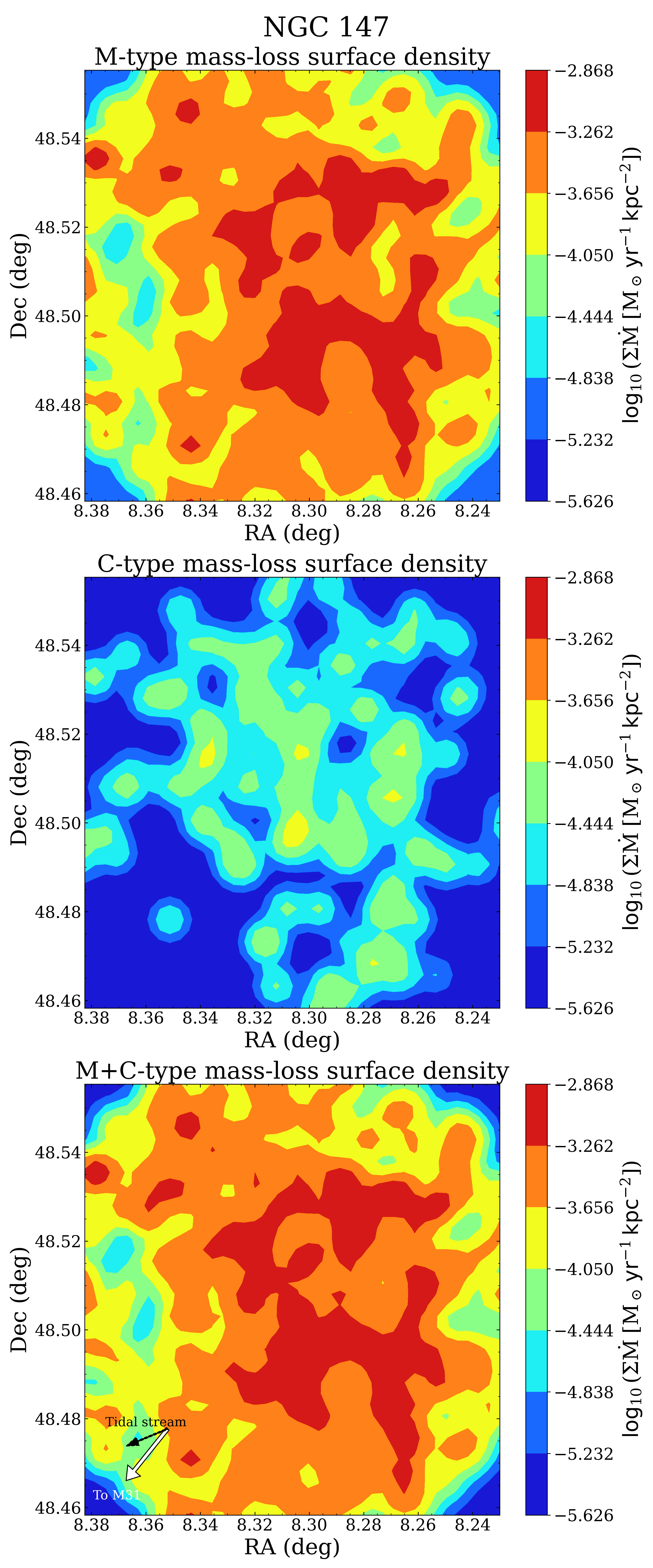}
\includegraphics[width=0.45\textwidth]{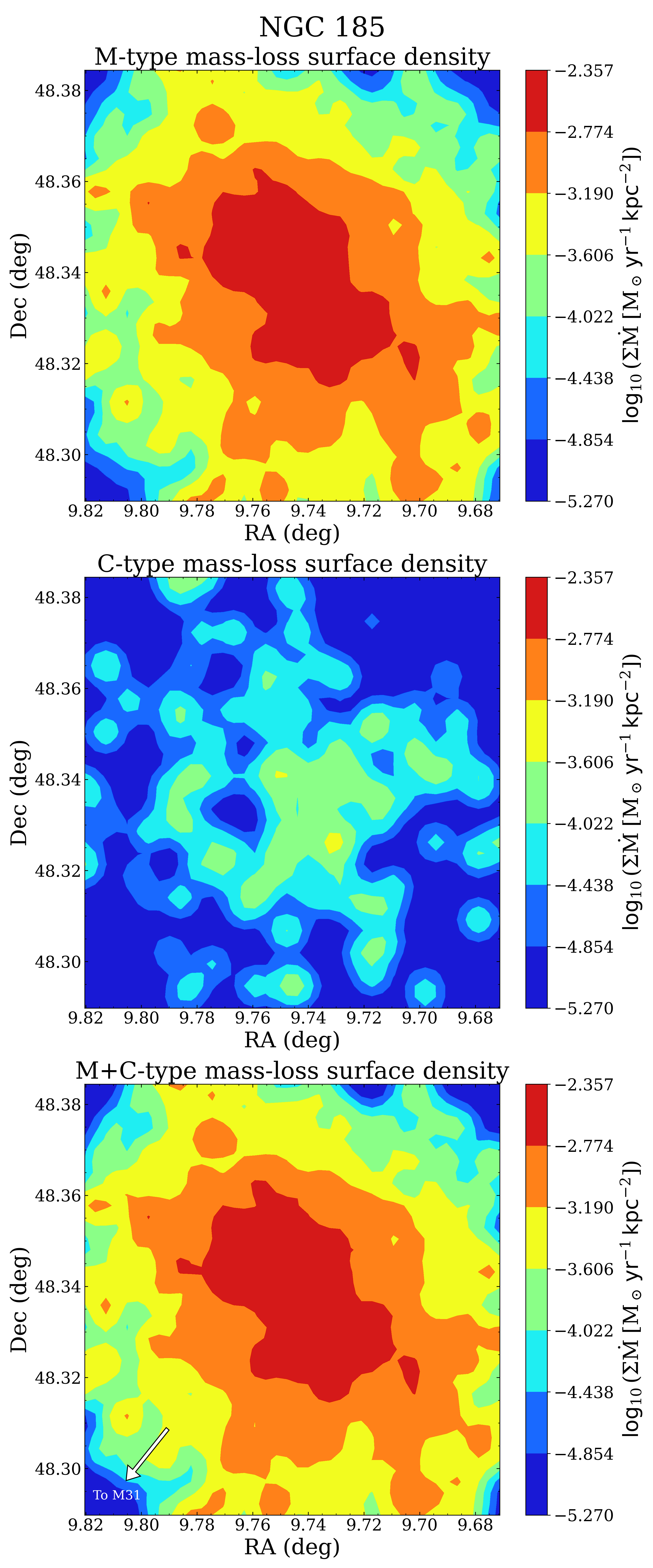}
\caption{ Mass-loss rate surface density maps for carbon-rich (C-type), oxygen-rich (M-type), and combined (M+C-type) stars in galaxies NGC 147 and NGC 185. White arrows indicate the approximate direction toward M31, and the black arrow in the NGC 147 panel shows the approximate direction of the tidal stream, following \citet{Preston25}.} 
\label{fig:Mdot_Contour}
\end{center}
\end{figure*}

The structural disparity between NGC 185 and NGC 147 is further supported by the mass-loss rate surface density distribution, as depicted in Figure \ref{fig:Mdot_Contour}, which illustrates the mass-loss rate surface density in $\msun yr^{-1} kpc^{-2}$ for these two galaxies. The patterns observed in the mass-loss rate surface density  plots mirror those discussed above, with NGC 185 exhibiting a centralized concentration and NGC 147 showing a more fragmented distribution. To enhance visualization, the number density plots are scaled using the square root to better highlight variations in stellar density, while the mass-loss rate surface density plots employ a logarithmic scale to effectively represent the wide range of mass-loss rates across the galaxies.

\begin{table}[ht]
\centering
\caption{Spatial properties of AGB stars}
\label{tab:spatial_stats}
\begin{tabular}{lcc}
\hline
\hline
 & \textbf{NGC 185} & \textbf{NGC 147} \\
\hline
\textit{M-type stars} & & \\
\hspace{2mm} Count ($N$) & 1732 & 950 \\
\hspace{2mm} RA, Dec (deg) & 9.74, 48.33 & 8.30, 48.51 \\
\hspace{2mm} Dispersion (arcsec) & $64.54 \pm 1.10$ & $65.63 \pm 1.51$ \\
\hspace{2mm} K-S $D$, $P$ & 0.17, 0.00 & 0.17, 0.00 \\
\\[-1ex]
\textit{C-type stars} & & \\
\hspace{2mm} Count ($N$) & 154 & 146 \\
\hspace{2mm} RA, Dec (deg) & 9.74, 48.33 & 8.30, 48.51 \\
\hspace{2mm} Dispersion (arcsec) & $67.34 \pm 3.84$ & $68.11 \pm 3.99$ \\
\hspace{2mm} K-S $D$, $P$ & 0.10, 0.40 & 0.10, 0.40 \\
\hline
\hline
\end{tabular}
\end{table}

%@@@@@@@@@@@@@@@@@@@@@@@@@@@@@@@@@@@@@@@@@@@@@@@@@@@@@@@@@@@@@@@@@@@@@@@@@@@@@@@@@@@@@@@@@@

\section{Conclusions}
\label{sec:conclusion}

By fitting the spectral energy distribution to the data of long period variable stars at different wavelengths, we obtained their mass-loss rates in two dwarf galaxies, NGC 147 and NGC 185. The results of this study provide significant insights into the stellar mass-loss processes in these galaxies.

\begin{itemize}

% \item The total mass-loss rate of LPV stars in NGC 147 is estimated to be $(7.8 \pm 3.10) \times 10^{-5}$, while for NGC 185, it is $(3.0 \pm 1.17) \times 10^{-4}$. When calculating the maximum possible mass-loss rates without considering the chemical compositions, these values increase to $2.5 \times 10^{-4} \msun yr^{-1}$ and  $2.2 \times 10^{-4} \msun yr^{-1}$, respectively. For AGB stars, the total mass-loss rate is estimated at $9.0 \times 10^{-3} \msun yr^{-1}$ in NGC 147 and  $1.0 \times 10^{-2} \msun yr^{-1}$ in NGC 185.

% \item The mass-loss rate of evolved stars, while significant, proves insufficient to fully account for the mass budget of the target galaxies. Additional sources and mechanisms of mass-loss or mass influx need to be considered to account for the observed mass discrepancies.

% \item A two-dimensional map of dust dispersion reveals the gravitational interaction between NGC 147 and the Andromeda galaxy. This interaction likely influences the distribution and dynamics of the interstellar dust in NGC 147.

%\item By analyzing the dispersion or segregation of oxygen and carbon stars in the parametric space, we can gain insights into the star formation activity and quenching processes within the galaxy. The distribution of these stars serves as an indicator of the galaxy's current state, whether it is actively forming stars or in a quiescent phase.

\item The total mass-loss rate from LPV stars is estimated to be $(7.8 \pm 3.1) \times 10^{-5}\ \mathrm{M}_\odot\,\mathrm{yr}^{-1}$ in NGC 147 and $(3.0 \pm 1.2) \times 10^{-4}\ \mathrm{M}_\odot\,\mathrm{yr}^{-1}$ in NGC 185. 

\item For all reported AGB stars, the total mass-loss rates are $(9.4 \pm 3.8) \times 10^{-4}\ \mathrm{M}_\odot\,\mathrm{yr}^{-1}$ in NGC 147 and $(1.6 \pm 0.6) \times 10^{-3}\ \mathrm{M}_\odot\,\mathrm{yr}^{-1}$ in NGC 185, corresponding to dust mass injection rates of $(5.9 \pm 2.4) \times 10^{-6}\ \mathrm{M}_\odot\,\mathrm{yr}^{-1}$ and $(9.9 \pm 4.0) \times 10^{-6}\ \mathrm{M}_\odot\,\mathrm{yr}^{-1}$, respectively. 

%\item The specific mass return rates are $8.1 \times 10^{-12}\ \mathrm{yr}^{-1}$ for NGC 147 and $6.5 \times 10^{-11}\ \mathrm{yr}^{-1}$ for NGC 185, indicating that the mass returned by AGB stars is insufficient to replenish the current stellar mass. 

\item A two-dimensional map of dust dispersion suggests that gravitational interactions between NGC 147 and the Andromeda galaxy influence the distribution and dynamics of interstellar dust in NGC 147.\mbox{}\\

\end{itemize}

\section*{Acknowledgments}

H.M. would like to extend his sincere gratitude to Steven R. Goldman for his invaluable comments and assistance in using the \texttt{DESK} code. His expertise and insights have significantly contributed to the advancement of this research. Additionally, H.M. express his heartfelt thanks to ICE-CSIC for their generous hospitality at the Institute de Ciències de l'Espai and for the engaging and fruitful discussions that have greatly enriched this study.

%@@@@@@@@@@@@@@@@@@@@@@@@@@@@@@@@@@@@@@@@@@@@@@@@@@@@@@@@@@@@@@@@@@@@@@@@@@@@@@@@@@@@@@@@@@

\bibliography{NGCs}{}
\bibliographystyle{aasjournal}

%@@@@@@@@@@@@@@@@@@@@@@@@@@@@@@@@@@@@@@@@@@@@@@@@@@@@@@@@@@@@@@@@@@@@@@@@@@@@@@@@@@@@@@@@@@

\appendix
\section{Supplementary Material}
\label{sec:apndix}

\renewcommand{\thefigure}{A\arabic{figure}}
\renewcommand{\thetable}{A\arabic{table}}
\setcounter{figure}{0} % Reset the figure counter
\setcounter{table}{0}  % Reset the table counter

The final LPV catalogs for the galaxies NGC 147 and NGC 185, along with results and the best-fitted SEDs across different bands for these stars, are provided below.

%@@@@@@@@@@@@@@@@@@@@@@@@@@@@@@@@@@@@@@@@@@@@@@@@@@@@@@@@@@@@@@@@@@@@@@@@@@@@@@@@@@@@@@@@@@

\setlength{\LTcapwidth}{\textwidth}
%\clearpage
%\onecolumn
{\small % Adjust the font size to make the table smaller
\setlength{\tabcolsep}{5.2pt} % Adjust column spacing
\begin{longtable*}{@{}ccccccccccccccc@{}}
\caption {NGC 147 magnitudes. The ID column lists the unique identifier for each star in our final catalog. The RA, DEC, IRAC1, and IRAC2 columns are sourced from \citet{Boyer15a} catalog. Columns I and  Ks are from the \citet{Lorenz11} catalog, columns K, J$-$K, and H$-$K are from the \citet{Sohn06} catalog, columns i and V$-$i are from the \citet{Nowotny03} catalog, and columns I2 and R$-$I are from the \citet{Battinelli147} catalog.}
\label{tab:147mags}\\
\hline
\hline
\textbf{ID}	&	\textbf{RA}	&	\textbf{DEC}	&	\textbf{IRAC1}	&	\textbf{IRAC2}&	\textbf{I}	&	\textbf{Ks}	&	\textbf{K}	&		\textbf{J$-$K} &	\textbf{H$-$K}		&	\textbf{i}	&	\textbf{V$-$i}	&	\textbf{I2}	&		\textbf{R$-$I}	\\

 & (J2000) & (J2000) & [mag] & [mag]& [mag]& [mag]& [mag]& [mag]& [mag]& [mag]& [mag]
 & [mag]& [mag] \\ 
\hline
\endfirsthead

\multicolumn{15}{c}%
{{\tablename\ \thetable{} -- continued from previous page}} \\
\hline
\hline
\textbf{ID}	&	\textbf{RA}	&	\textbf{DEC}	&	\textbf{IRAC1}	&	\textbf{IRAC2}&	\textbf{I}	&	\textbf{Ks}	&	\textbf{K}	&		\textbf{J$-$K} &	\textbf{H$-$K}		&	\textbf{i}	&	\textbf{V$-$i}	&	\textbf{I2}	&		\textbf{R$-$I}	\\
 & (J2000) & (J2000) & [mag] & [mag]& [mag]& [mag]& [mag]& [mag]& [mag]& [mag]& [mag]
 & [mag]& [mag] \\ 
\hline
\endhead

\hline \multicolumn{15}{|r|}{{Continued on next page}} \\ \hline
\endfoot

\hline
\endlastfoot

1	&	00 33 24.32	&	+48 31 36.5	&	16.43	&	16.51	&	99.999	&	99.999	&	16.846	&	1.669	&	0.669	&	99.999	&	99.999	&	99.999	&	99.999	&	\\
2	&	00 33 24.11	&	+48 31 32.2	&	15.95	&	15.41	&	99.999	&	99.999	&	16.913	&	2.168	&	1.005	&	99.999	&	99.999	&	99.999	&	99.999	&	\\
3	&	00 33 22.37	&	+48 31 08.7	&	14.96	&	14.22	&	99.999	&	99.999	&	17.098	&	3.137	&	1.847	&	99.999	&	99.999	&	99.999	&	99.999	&	\\
4	&	00 33 21.79	&	+48 30 44.3	&	15.74	&	15.85	&	99.999	&	99.999	&	16.313	&	1.545	&	0.531	&	99.999	&	99.999	&	99.999	&	99.999	&	\\
5	&	00 33 21.80	&	+48 31 24.0	&	15.94	&	16.13	&	99.999	&	99.999	&	16.883	&	1.683	&	0.694	&	99.999	&	99.999	&	99.999	&	99.999	&	\\
6	&	00 33 21.60	&	+48 30 03.8	&16.00	&	16.14	&	99.999	&	99.999	&	16.86	&	1.941	&	0.787	&	99.999	&	99.999	&	99.999	&	99.999	&	\\
7	&	00 33 19.56	&	+48 30 51.7	&	15.12	&	14.53	&	99.999	&	99.999	&	16.758	&	3.411	&	1.527	&	99.999	&	99.999	&	99.999	&	99.999	&	\\
8	&	00 33 18.36	&	+48 29 46.7	&	16.11	&	16.24	&	99.999	&	99.999	&	16.679	&	1.406	&	0.467	&	99.999	&	99.999	&	99.999	&	99.999	&	\\
\hline
\end{longtable*}
}

%@@@@@@@@@@@@@@@@@@@@@@@@@@@@@@@@@@@@@@@@@@@@@@@@@@@@@@@@@@@@@@@@@@@@@@@@@@@@@@@@@@@@@@@@@@@@
\clearpage

\begin{longtable}{@{}cccccccccccc@{}}
\caption{NGC 147 results. The ID column provides the unique identifier for each star in our final catalog. The subscripts M and C denote the oxygen or carbon model used, respectively. $\tau$ represents the optical depth, L indicates the luminosity, and $ \dot{M} $ is the mass-loss rate. BC stands for the bolometric correction. $M_{ini}$ refers to the initial mass value derived from the calculations in Section \ref{sec:code}. The Chemical type column specifies the star's chemical composition as reported in the literature, while the Mass type column expresses this composition in terms of the initial mass.} \label{tab:147results} \\
\hline
\hline
\textbf{ID} & \boldmath$\tau_{C}$ & \boldmath$\tau_{M}$ & \boldmath$L_{C}$ & \boldmath$L_{M}$ & \boldmath$\dot{M}_{C}$ & \boldmath$\dot{M}_{M}$ & \boldmath$BC_{C}$ & \boldmath$BC_{M}$ & \boldmath$M_{ini}$ & \textbf{Chemical type} & \textbf{Mass type} \\
 &  &  & [\lsun] & [\lsun]& [\msun/yr] & [\msun/yr] & [mag] & [mag]& [\msun] & &  \\
\hline
\endfirsthead

\multicolumn{12}{c}%
{{\tablename\ \thetable{} -- continued from previous page}} \\
\hline
\hline
\textbf{ID} & \boldmath$\tau_{C}$ & \boldmath$\tau_{M}$ & \boldmath$L_{C}$ & \boldmath$L_{M}$ & \boldmath$\dot{M}_{C}$ & \boldmath$\dot{M}_{M}$ & \boldmath$BC_{C}$ & \boldmath$BC_{M}$ & \boldmath$M_{ini}$ & \textbf{Chemical type} & \textbf{Mass type} \\
 &  &  & [\lsun] & [\lsun]& [\msun/yr] & [\msun/yr] & [mag] & [mag]& [\msun] & &  \\
\hline
\endhead

\hline \multicolumn{12}{|r|}{{Continued on next page}} \\ \hline
\endfoot

\hline
\endlastfoot

1	&	0.01	&	0.01	&	3550	&	3400	&	2.07E-08	&	1.87E-07	&	3.23	&	3.28	&	1.07	&	C	&	M	\\
2	&	0.2	&	0.41	&	3510	&	4130	&	1.17E-06	&	5.66E-06	&	3.17	&	3	&	1.11	&	C	&	C	\\
3	&	0.95	&	1.25	&	5410	&	7640	&	4.13E-06	&	2.53E-05	&	2.52	&	2.14	&	1.01	&	C	&	M	\\
4	&	0.01	&	0.01	&	6220	&	6420	&	3.03E-08	&	3.01E-07	&	3.15	&	3.12	&	1.73	&	C	&	C	\\
5	&	0.02	&	0.02	&	3370	&	3390	&	3.36E-08	&	1.3E-07	&	3.25	&	3.24	&	1.18	&	C	&	C	\\
6	&	0.06	&	0.01	&	3180	&	3170	&	9.65E-08	&	5.74E-08	&	3.33	&	3.34	&	1.15	&	C	&	C	\\
7	&	0.85	&	1.15	&	4540	&	5650	&	2.16E-06	&	1.31E-05	&	3.05	&	2.81	&	1.31	&	C	&	C	\\
8	&	0.01	&	0.01	&	4770	&	4940	&	2.58E-08	&	2.62E-07	&	3.08	&	3.04	&	1.35	&	C	&	C	\\
\hline

\end{longtable}

%@@@@@@@@@@@@@@@@@@@@@@@@@@@@@@@@@@@@@@@@@@@@@@@@@@@@@@@@@@@@@@@@@@@@@@@@@@@@@@@@@@@@@@@@@@@@
\clearpage

{\small % Adjust the font size to make the table smaller
\setlength{\tabcolsep}{5.2pt} % Adjust column spacing
\begin{longtable}{@{}ccccccccccccccc@{}}
\caption{NGC 185 magnitudes. The ID column lists the unique identifier for each star in our final catalog. The RA, DEC, IRAC1, and IRAC2 columns are sourced from \citet{Boyer15a} catalog. Columns I and  Ks are from the \citet{Lorenz11} catalog, columns J, J$-$K, and H$-$K are from the \citet{Kang05} catalog, columns i and V$-$i are from the \citet{Nowotny03} catalog, and columns I2 and R$-$I are from the \citet{Battinelli185} catalog.} \label{tab:185mags} \\
\hline
\hline
\textbf{ID}	&	\textbf{RA}	&	\textbf{DEC}	&	\textbf{IRAC1}	&	\textbf{IRAC2}&	\textbf{I}	&	\textbf{Ks}	&	\textbf{J}	&		\textbf{J$-$K} &	\textbf{H$-$K}		&	\textbf{i}	&	\textbf{V$-$i}	&	\textbf{I2}	&		\textbf{R$-$I}	\\

 & (J2000) & (J2000) & [mag] & [mag]& [mag]& [mag]& [mag]& [mag]& [mag]& [mag]& [mag]
 & [mag]& [mag] \\ 
\hline
\endfirsthead

\multicolumn{15}{c}%
{{\tablename\ \thetable{} -- continued from previous page}} \\
\hline
\hline
\textbf{ID}	&	\textbf{RA}	&	\textbf{DEC}	&	\textbf{IRAC1}	&	\textbf{IRAC2}&	\textbf{I}	&	\textbf{Ks}	&	\textbf{J}	&		\textbf{J$-$K} &	\textbf{H$-$K}		&	\textbf{i}	&	\textbf{V$-$i}	&	\textbf{I2}	&		\textbf{R$-$I}	\\
 & (J2000) & (J2000) & [mag] & [mag]& [mag]& [mag]& [mag]& [mag]& [mag]& [mag]& [mag]
 & [mag]& [mag] \\ 
\hline
\endhead

\hline \multicolumn{15}{|r|}{{Continued on next page}} \\ \hline
\endfoot

\hline
\endlastfoot

1	&	00 39 03.83	&	+48 19 16.1	&	15.08	&	14.67	&	99.999	&	99.999	&	18.594	&	2.466	&	1.04	&	99.999	&	99.999	&	99.999	&	99.999	\\
2	&	00 39 00.40	&	+48 21 30.8	&	15.18	&	14.54	&	99.999	&	99.999	&	19.48	&	3.092	&	1.203	&	99.999	&	99.999	&	99.999	&	99.999	\\
3	&	00 38 59.54	&	+48 21 41.4	&	14.82	&	14.33	&	99.999	&	99.999	&	19.43	&	3.102	&	1.237	&	99.999	&	99.999	&	99.999	&	99.999	\\
4	&	00 38 56.87	&	+48 21 17.8	&	15.22	&	15.3	&	99.999	&	99.999	&	17.748	&	2.053	&	0.694	&	99.999	&	99.999	&	99.999	&	99.999	\\
5	&	00 38 55.13	&	+48 20 04.8	&	15.52	&	15.77	&	99.999	&	99.999	&	19.02	&	2.731	&	0.869	&	99.999	&	99.999	&	99.999	&	99.999	\\
6	&	00 38 52.21	&	+48 21 46.7	&	14.96	&	14.41	&	99.999	&	99.999	&	19.843	&	3.522	&	1.358	&	99.999	&	99.999	&	99.999	&	99.999	\\
7	&	00 39 14.34	&	+48 18 37.3	&	17.85	&	17.75	&	19.84	&	99.999	&	99.999	&	99.999	&	99.999	&	19.88	&	1.8	&	99.999	&	99.999	\\
8	&	00 39 13.02	&	+48 19 25.5	&	17.16	&	17.11	&	19.57	&	17.48	&	99.999	&	99.999	&	99.999	&	19.26	&	1.91	&	99.999	&	99.999	\\
\hline

\end{longtable}
}

%@@@@@@@@@@@@@@@@@@@@@@@@@@@@@@@@@@@@@@@@@@@@@@@@@@@@@@@@@@@@@@@@@@@@@@@@@@@@@@@@@@@@@@@@@@@@
\clearpage

\begin{longtable}{@{}cccccccccccc@{}}
\caption{NGC 185 results. The ID column provides the unique identifier for each star in our final catalog. The subscripts M and C denote the oxygen or carbon model used, respectively. $\tau$ represents the optical depth, L indicates the luminosity, and $ \dot{M} $ is the mass-loss rate. BC stands for the bolometric correction. $M_{ini}$ refers to the initial mass value derived from the calculations in Section \ref{sec:code}. The Chemical type column specifies the star's chemical composition as reported in the literature, while the Mass type column expresses this composition in terms of the initial mass.} \label{tab:185results} \\
\hline
\hline
\textbf{ID} & \boldmath$\tau_{C}$ & \boldmath$\tau_{M}$ & \boldmath$L_{C}$ & \boldmath$L_{M}$ & \boldmath$\dot{M}_{C}$ & \boldmath$\dot{M}_{M}$ & \boldmath$BC_{C}$ & \boldmath$BC_{M}$ & \boldmath$M_{ini}$ & \textbf{Chemical type} & \textbf{Mass type} \\
 &  &  & [\lsun] & [\lsun]& [\msun/yr] & [\msun/yr] & [mag] & [mag]& [\msun] & &  \\
\hline
\endfirsthead

\multicolumn{12}{c}%
{{\tablename\ \thetable{} -- continued from previous page}} \\
\hline
\hline
\textbf{ID} & \boldmath$\tau_{C}$ & \boldmath$\tau_{M}$ & \boldmath$L_{C}$ & \boldmath$L_{M}$ & \boldmath$\dot{M}_{C}$ & \boldmath$\dot{M}_{M}$ & \boldmath$BC_{C}$ & \boldmath$BC_{M}$ & \boldmath$M_{ini}$ & \textbf{Chemical type} & \textbf{Mass type} \\
 &  &  & [\lsun] & [\lsun]& [\msun/yr] & [\msun/yr] & [mag] & [mag]& [\msun] & &  \\
\hline
\endhead

\hline \multicolumn{12}{|r|}{{Continued on next page}} \\ \hline
\endfoot

\hline
\endlastfoot

1	&	0.46	&	0.44	&	5080	&	5490	&	1.12E-06	&	3.57E-06	&	3.36	&	3.27	&	2.03	&	C	&	C	\\
2	&	0.7	&	0.95	&	4200	&	8580	&	1.25E-06	&	2.98E-05	&	3.3	&	2.53	&	1.76	&	C	&	C	\\
3	&	0.7	&	0.85	&	5060	&	6100	&	1.75E-06	&	8.76E-06	&	3.16	&	2.96	&	1.88	&	C	&	C	\\
4	&	0.02	&	0.01	&	7040	&	6780	&	6.13E-08	&	2.6E-07	&	3.44	&	3.48	&	2.24	&	C	&	C	\\
5	&	0.25	&	0.28	&	3520	&	3770	&	4.1E-07	&	1.75E-06	&	3.59	&	3.52	&	1.56	&	C	&	C	\\
6	&	0.9	&	1	&	4530	&	5410	&	1.67E-06	&	7.63E-06	&	3.29	&	3.1	&	1.79	&	C	&	C	\\
7	&	0.01	&	0.01	&	1940	&	1920	&	6.69E-08	&	1.93E-07	&	-	&	-	&	0.88	&	M	&	M	\\
8	&	0.01	&	0.01	&	3010	&	3160	&	2.31E-08	&	2.79E-07	&	-	&	-	&	0.96	&	M	&	M	\\
\hline

\end{longtable}

%@@@@@@@@@@@@@@@@@@@@@@@@@@@@@@@@@@@@@@@@@@@@@@@@@@@@@@@@@@@@@@@@@@@@@@@@@@@@@@@@@@@@@@@@@@@@
\clearpage

\begin{figure}
    \centering
    \includegraphics[width=0.9\textwidth]{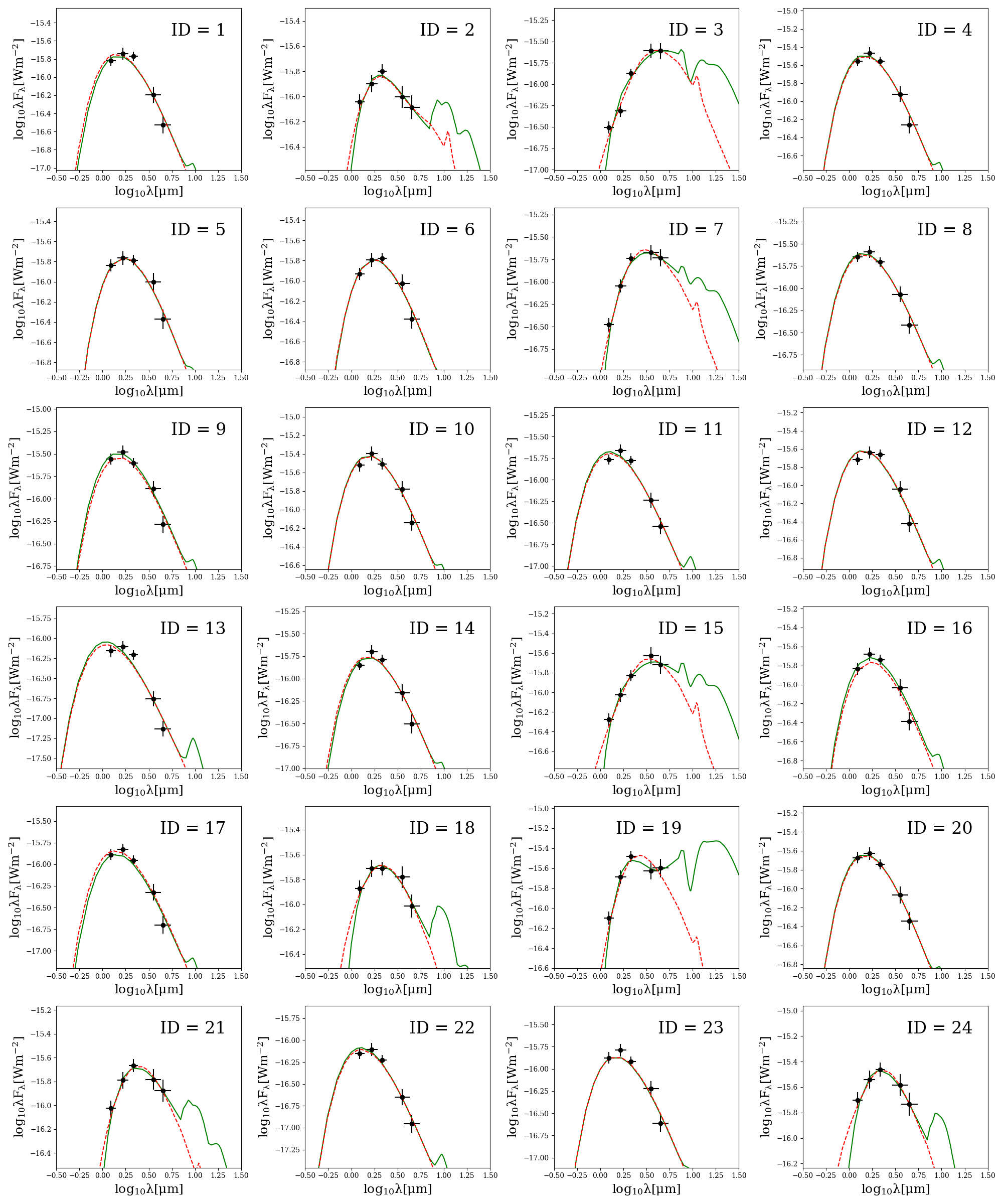}
    \begin{minipage}{0.85\textwidth} % Match the figure width
        \caption{Spectral energy distribution fits for LPV stars in the final NGC 147 catalog. The complete figure set (7 plots), is available in the online journal. Observed photometric data are shown as black circles with error bars, which represent the photometric uncertainties reported in the original catalogs (see Section \ref{sec:catalogs}). The solid green lines correspond to best-fitting oxygen-rich models, while the red dashed lines represent best-fitting carbon-rich models.}
        \label{fig:SED_fits_147}
    \end{minipage}
%    \figsetstart
%    \figsetnum{A1}
%    \figsettitle{SED Fits of LPV Stars in NGC 147}
%    \figsetgrpstart
    \figsetgrpnum{A1.1}
    \figsetgrptitle{SED Fit (a)}
    \figsetplot{147P1.png}
    \figsetgrpnote{Best-fit SED for LPV star (a) in NGC 147.}
    \figsetgrpend
    \figsetgrpstart
    \figsetgrpnum{A1.2}
    \figsetgrptitle{SED Fit (b)}
    \figsetplot{147P2.png}
    \figsetgrpnote{Best-fit SED for LPV star (b) in NGC 147.}
    \figsetgrpend
    \figsetgrpstart
    \figsetgrpnum{A1.3}
    \figsetgrptitle{SED Fit (c)}
    \figsetplot{147P3.png}
    \figsetgrpnote{Best-fit SED for LPV star (c) in NGC 147.}
    \figsetgrpend
    \figsetgrpstart
    \figsetgrpnum{A1.4}
    \figsetgrptitle{SED Fit (d)}
    \figsetplot{147P4.png}
    \figsetgrpnote{Best-fit SED for LPV star (d) in NGC 147.}
    \figsetgrpend
    \figsetgrpstart
    \figsetgrpnum{A1.5}
    \figsetgrptitle{SED Fit (e)}
    \figsetplot{147P5.png}
    \figsetgrpnote{Best-fit SED for LPV star (e) in NGC 147.}
    \figsetgrpend
    \figsetgrpstart
    \figsetgrpnum{A1.6}
    \figsetgrptitle{SED Fit (f)}
    \figsetplot{147P6.png}
    \figsetgrpnote{Best-fit SED for LPV star (f) in NGC 147.}
    \figsetgrpend
    \figsetgrpstart
    \figsetgrpnum{A1.7}
    \figsetgrptitle{SED Fit (g)}
    \figsetplot{147P7.png}
    \figsetgrpnote{Best-fit SED for LPV star (g) in NGC 147.}
    \figsetgrpend
    \figsetend
\end{figure}
\clearpage % Ensure all floats are flushed

%@@@@@@@@@@@@@@@@@@@@@@@@@@@@@@@@@@@@@@@@@@@@@@@@@@@@@@@@@@@@@@@@@@@@@@@@@@@@@@@@@@@@@@@@@@@@

%@@@@@@@@@@@@@@@@@@@@@@@@@@@@@@@@@@@@@@@@@@@@@@@@@@@@@@@@@@@@@@@@@@@@@@@@@@@@@@@@@@@@@@@@@@@@
\clearpage

\begin{figure}
    \centering
    \includegraphics[width=0.9\textwidth]{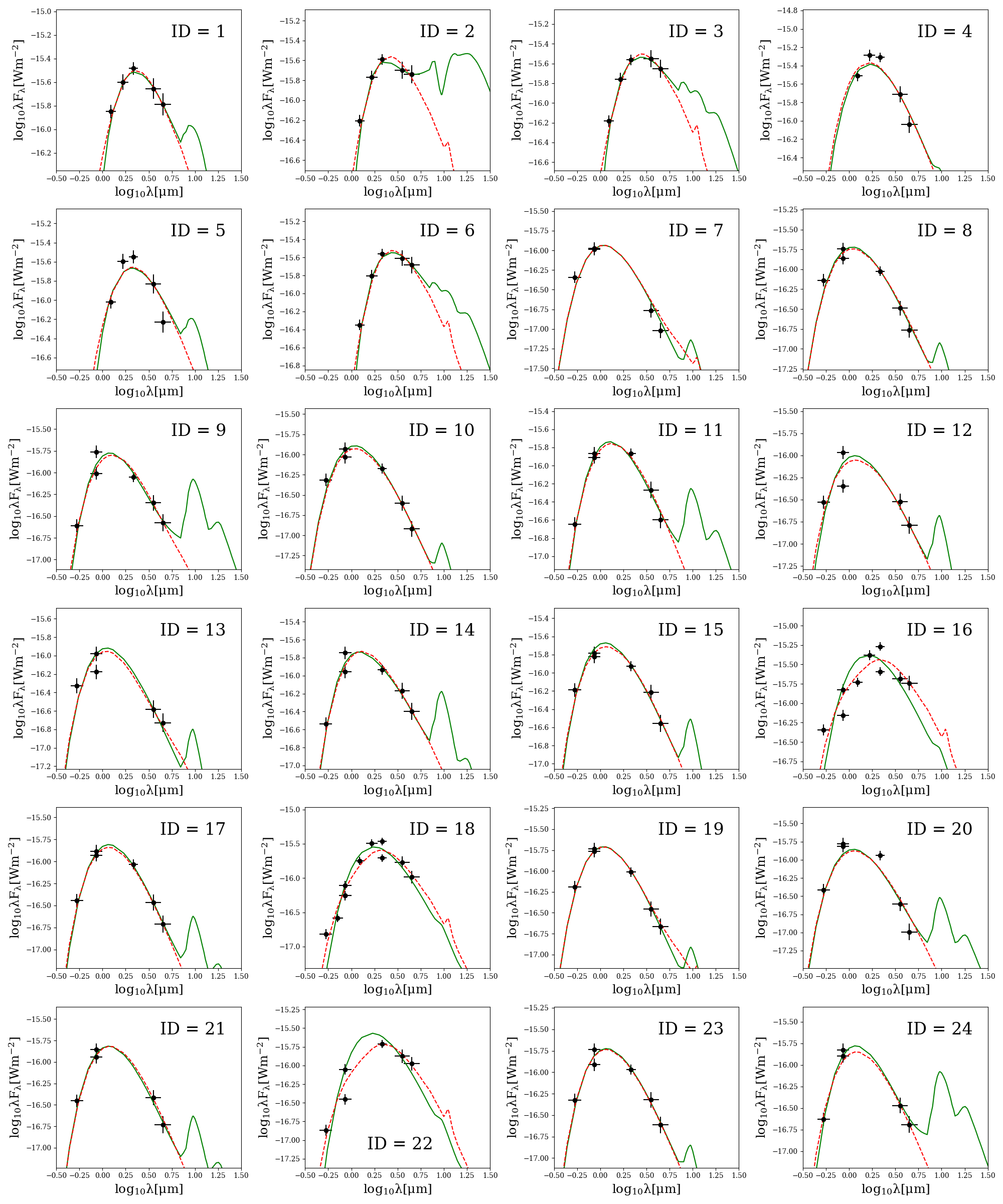}
    \begin{minipage}{0.85\textwidth} % Match the figure width
        \caption{Similar to Figure \ref{fig:SED_fits_147}, but for NGC 185. The complete figure set, consisting of 8 plots, is available in the online journal.}
        \label{fig:SED_fits_185}
    \end{minipage}
%    \figsetstart
%    \figsetnum{A2}
%    \figsettitle{SED Fits of LPV Stars in NGC 185}
%    \figsetgrpstart
    \figsetgrpnum{A2.1}
    \figsetgrptitle{SED Fit (a)}
    \figsetplot{185P1.png}
    \figsetgrpnote{Best-fit SED for LPV star (a) in NGC 185.}
    \figsetgrpend
    \figsetgrpstart
    \figsetgrpnum{A2.2}
    \figsetgrptitle{SED Fit (b)}
    \figsetplot{185P2.png}
    \figsetgrpnote{Best-fit SED for LPV star (b) in NGC 185.}
    \figsetgrpend
    \figsetgrpstart
    \figsetgrpnum{A2.3}
    \figsetgrptitle{SED Fit (c)}
    \figsetplot{185P3.png}
    \figsetgrpnote{Best-fit SED for LPV star (c) in NGC 185.}
    \figsetgrpend
    \figsetgrpstart
    \figsetgrpnum{A2.4}
    \figsetgrptitle{SED Fit (d)}
    \figsetplot{185P4.png}
    \figsetgrpnote{Best-fit SED for LPV star (d) in NGC 185.}
    \figsetgrpend
    \figsetgrpstart
    \figsetgrpnum{A2.5}
    \figsetgrptitle{SED Fit (e)}
    \figsetplot{185P5.png}
    \figsetgrpnote{Best-fit SED for LPV star (e) in NGC 185.}
    \figsetgrpend
    \figsetgrpstart
    \figsetgrpnum{A2.6}
    \figsetgrptitle{SED Fit (f)}
    \figsetplot{185P6.png}
    \figsetgrpnote{Best-fit SED for LPV star (f) in NGC 185.}
    \figsetgrpend
    \figsetgrpstart
    \figsetgrpnum{A2.7}
    \figsetgrptitle{SED Fit (g)}
    \figsetplot{185P7.png}
    \figsetgrpnote{Best-fit SED for LPV star (g) in NGC 185.}
    \figsetgrpend
    \figsetgrpstart
    \figsetgrpnum{A2.8}
    \figsetgrptitle{SED Fit (h)}
    \figsetplot{185P8.png}
    \figsetgrpnote{Best-fit SED for LPV star (h) in NGC 185.}
    \figsetgrpend
    \figsetend
\end{figure}
\clearpage % Ensure all floats are flushed

%@@@@@@@@@@@@@@@@@@@@@@@@@@@@@@@@@@@@@@@@@@@@@@@@@@@@@@@@@@@@@@@@@@@@@@@@@@@@@@@@@@@@@@@@@@@@

\end{document}